# Online Monotone Games

**Ian Gemp** and **Sridhar Mahadevan**
College of Information and Computer Sciences
University of Massachusetts Amherst
Amherst, MA 01003
{imgemp,mahadeva}@cics.umass.edu

## Abstract

Algorithmic game theory (AGT) focuses on the design and analysis of algorithms for interacting agents, with interactions rigorously formalized within the framework of games. Results from AGT find applications in domains such as online bidding auctions for web advertisements and network routing protocols. *Monotone* games are games where agent strategies naturally converge to an equilibrium state. Previous results in AGT have been obtained for convex, socially-convex, or smooth games, but not monotone games. Our primary theoretical contributions are defining the monotone game setting and its extension to the online setting, a new notion of regret for this setting, and accompanying algorithms that achieve sub-linear regret. We demonstrate the utility of online monotone game theory on a variety of problem domains including variational inequalities, reinforcement learning, and generative adversarial networks.

## 1 Introduction

Algorithmic game theory (AGT) focuses on the design and analysis of algorithms for agents interacting in games (Nisan et al. 2007). A game is formalized as follows. At each time step, $t$, each agent $i$ selects a strategy, $x^{(i)}$, from a set, $\mathcal{X}^{(i)}$, and incurs a cost, $f_t^{(i)}(x)$, where $x$ contains all players' selected strategies. The *welfare* of a strategy set is the sum of all player payoffs, i.e., $W_t(x) := -\sum_i f_t^{(i)}(x)$. Examples of games include auctions where bots bid for advertising space on a webpage and resource allocation games where peers vie for bandwidth on a network. AGT is typically concerned with one of two goals: 1) designing sophisticated multi-agent algorithms that achieve some desired global behavior (e.g., maximal welfare) or 2) examining the effect of simple, selfish agents on global behaviors. The former is most useful when one has central control over all the agent algorithms, while the latter is useful when analyzing more *natural* agent algorithms such as gradient descent. In either case, to make these goals tractable, the design and analysis is restricted to games with properties that algorithms can exploit.

In this work, we consider *repeated-games*, wherein agents have the opportunity to learn and extract information about the game by playing the game repeatedly. Agents begin play with no information about the game, and so they cannot be expected to play optimally from the start. We therefore analyze the *regret* of an agent (or algorithm), which measures how poorly an agent performs relative to some baseline. Algorithms that asymptotically achieve an average regret of zero are called *no-regret* algorithms.

Several types of repeated-games have been studied, including convex, socially-convex, and smooth games. Here, we consider a new type of game: *monotone games*. Monotone games are games where agent strategies naturally converge to an equilibrium state. Mathematically, the dynamics of monotone games satisfy a monotonicity property analogous to convexity in optimization, which, given convexity's importance for optimization, lends credence to monotonicity being a fundamental property of games.

In this work we illuminate the boundaries between monotone games and other well known types of games. We introduce a new performance metric, *auto*-welfare. We then show that monotone games are a subset of convex games, that they facilitate no-regret algorithms for *auto*-welfare, and that they need not be socially-convex or smooth. In addition, we show that *auto*-welfare is guaranteed even if the rules of the game change at every time step. To show a few of the implications of these results for real applications, we perform an online analysis of the family of GTD algorithms (Sutton, Maei, and Szepesvári 2009) for reinforcement learning and derive an affine Wasserstein generative adversarial network (WGAN) (Arjovsky, Chintala, and Bottou 2017). In summary, our primary contributions are

- the concept of (online) monotone games,
- the concept of *auto*-welfare regret, with accompanying linear bounds,
- algorithms that achieve sublinear *auto*-welfare regret,
- and examples of a variety of monotone games of interest.

## 2 Technical Background

Let $\tilde{x}^{(i)}$ be the vector of player strategies where player $i$ plays any $x^{(i)} \in \mathcal{X}^{(i)}$ and player $j \neq i$ plays $x^{*(j)}$. Then the vector of agent strategies, $x^* \in \mathcal{X}$, is a Nash equilibrium if for all $i$ and all $x^{(i)} \in \mathcal{X}^{(i)}$, $f^{(i)}(\tilde{x}^{(i)}) \geq f^{(i)}(x^*)$. We denote the set of all Nash equilibria by $\mathcal{X}^*$.

We denote the norm of a vector, $x$, by $||x||$, and its dual norm by $||x||_* = \max\{\langle x', x \rangle : ||x'|| \leq 1\}$. A set is *convex* if for all $x \in \mathcal{X}$, all $x' \in \mathcal{X}$, and all $t \in [0, 1]$, $tx + (1-t)x' \in$

$\mathcal{X}$. Unless stated otherwise, all sets are assumed to be convex subsets of $\mathbb{R}^n$.

The subdifferential of a convex function, $f : \mathcal{X} \to \mathbb{R}$, at $x$, denoted $\partial f(x)$, is the set of all subgradients at $x$, i.e., $\partial f(x) = \{z : \forall x' \in \mathcal{X}, \langle z, x' - x \rangle \leq f(x') - f(x)\}$. A function, $f$, is $L$-Lipschitz over $\mathcal{X}$ if $|f(x') - f(x)| \leq L||x'-x||$, or equivalently, if for all $x \in \mathcal{X}$ and all $z \in \partial f(x)$, $||z||_* \leq L$. A function, $f$, is *convex* if

$$\forall x \in \mathcal{X}, x' \in \mathcal{X}, z \in \partial f(x), z' \in \partial f(x'),$$
$$\langle z - z', x - x' \rangle \geq 0.^1 \qquad (1)$$

Similarly to $\partial f$, a set valued map, $F : \mathcal{X} \to \{\mathcal{Z} : \mathcal{Z} \subseteq \mathbb{R}^n\}$, maps elements of $\mathcal{X}$ to sets of elements in $\mathbb{R}^n$. The set, $\mathcal{Z}$, need not be convex. A map, $F$, is $L$-Lipschitz over $\mathcal{X}$ if for all $x \in \mathcal{X}$ and all $x' \in \mathcal{X}$, $||F(x') - F(x)|| \leq L||x'-x||$, or equivalently, if for all $x \in \mathcal{X}$ and all $z \in F(x)$, $||z||_* \leq L$. The definition of convexity can be extended to apply to maps, rather than real-valued functions. This generalization of convexity is called *monotonicity*. Formally, a map, $F$, is *monotone* (Aslam Noor 1998) if

$$\forall x \in \mathcal{X}, x' \in \mathcal{X}, z \in F(x), z' \in F(x'),$$
$$\langle z - z', x - x' \rangle \geq 0. \qquad (2)$$

Alternatively, if $F$ is differentiable and the Jacobian of $F$ is positive semidefinite (PSD), then $F$ is monotone (Nagurney and Zhang 1996; Schaible and Luc 1996). A (possibly nonsymmetric) matrix, $A$, is PSD if for all $x \in \mathbb{R}^n$, $x^T A x \geq 0$, or equivalently, $A$ is PSD if $(A+A^T) \succeq 0$. Although an abuse of notation, we sometimes write $F$ or $\partial f$ to represent elements from their respective sets as opposed to the sets themselves—their use should be clear from the context.

The concept of a *path integral* is central to this work. A path (also *line* or *contour*) integral is an integral along a directed, scalar-parameterized path, $x : \mathbb{R} \to \mathcal{X}$, from $a \in \mathcal{X}$ to $b \in \mathcal{X}$, of the inner product between an output of the map and a differential of the curve, $dx$, i.e.,

$$\int_{x:a\leadsto b} \langle F(x), dx \rangle = \int_0^1 \langle F(x(t)), \frac{dx}{dt} \rangle dt \qquad (3)$$

where $x(0) = a$ and $x(1) = b$. By our definitions, $\langle F(x), dx \rangle$ is guaranteed to be integrable over finite length paths through the domain $\mathcal{X}$. A path is closed if $a = b$. $F$ is *conservative* if the integral along any closed path in $\mathcal{X}$ is zero, i.e., $\oint \langle F(x), dx \rangle = 0$.[2] Equivalently, $F$ is conservative if it is path-independent, meaning the integral along any path between two points in $\mathcal{X}$ does not depend on the path itself. Conversely, if $F$ is path-dependent or $\exists x \in \mathcal{X}$ such that $\oint \langle F(x), dx \rangle \neq 0$, then $F$ is not conservative.

By (1) and (2), the subdifferential of a convex function is a monotone set valued map, but the converse is not necessarily true. If $F$ is monotone and conservative, then $F$ is the subdifferential of some convex function (Romano et al. 1993), however, not all monotone maps are conservative. This shows that monotone maps are capable of representing problems not readily captured by convex functions. A more thorough discussion relating monotone maps to subdifferentials can be found in the work of Romano et al., (1993).

---
[1] Convexity is typically presented as $f(tx+(1-t)x') \leq tf(x) + (1-t)f(x')$, but the form we present better suits the comparison to monotonicity.

[2] This is colloquially known as the *fundamental theorem of calculus for line integrals*.

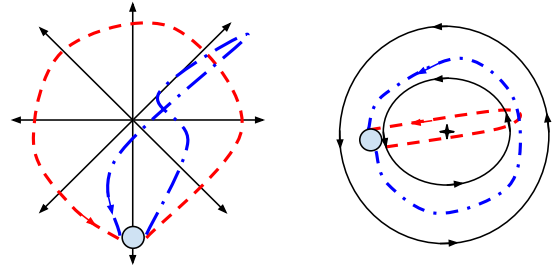

Figure 1: The sketch on the left represents the gradient map of a convex function (e.g., $x^2 + y^2$), while the sketch on the right is of a circular vector field and not the gradient of any function. Both are monotone maps, however, the blue and red contour integrals in the left sketch are over a conservative map and so they both evaluate to zero (path-independent). In contrast, the blue and red contour integrals in the right sketch are over a non-conservative monotone map and possibly evaluate to different values (path-dependent).

In Section 5.2 we apply our framework to a variational inequality (VI) problem, so we introduce VIs here. The variational inequality problem, $VI(F, \mathcal{X})$, is to find an $x^*$ such that for all $x \in \mathcal{X}$, $\langle F(x^*), x - x^* \rangle \geq 0$. VIs are used to model equilibrium problems in a number of domains including mechanics, traffic networks, economics, and game theory (Facchinei and Pang 2003). Under mild conditions (see Appendix A.3), $x^*$ constitutes a Nash equilibrium point. It is known that the solution set, $\mathcal{X}^*$, to $VI(F, \mathcal{X})$ with monotone $F$ is a convex, compact set (Crouzeix 2016). We refer the reader to the work of Rockafellar (1969) for a detailed study of the relationship between convex functions, monotone operators, and variational inequalities and to that of Nagurney and Zhang (1999; 1996) and Facchinei and Pang (2003) for an extensive study of VIs. Geometrically, monotonicity implies that $\mathcal{X}^*$ is a global monotone attractor (Nagurney and Zhang 1996). Figure 1 demonstrates a few of the aforementioned properties of monotonicity.

Online convex optimization (OCO) is a framework for studying the online learning problem when the loss, $f_t(x)$, is convex with respect to the prediction domain, $\mathcal{X}$, which is also a convex set. The learning problem is defined in Algorithm 1. We refer the reader to the survey by Shalev-Shwartz (2011) for a review of OCO. The standard no-regret algorithm for OCO is online gradient (mirror) descent; online gradient descent (OGD) is described in Algorithm 2. The regret of an online learning algorithm, $\mathcal{A}$, predicting sequence,

## Algorithm 1 Online Convex Optimization (OCO)

input: A convex set $\mathcal{X}$
**for all** $t = 1, 2, \ldots$ **do**
   predict a vector $x_t \in \mathcal{X}$
   receive a convex loss function $f_t : \mathcal{X} \to \mathbb{R}$
   suffer loss $f_t(x_t)$
**end for**

## Algorithm 2 Online Gradient Descent (OGD)

input: A scalar learning rate $\eta > 0$
$x_1 = 0$
**for all** $t = 1, 2, \ldots$ **do**
   $x_{t+1} = x_t - \eta z_t$ where $z_t \in \partial f_t(x_t)$
**end for**

$x_t$, after $T$ steps, is formally defined as

$$regret_{\mathcal{A}}^T(\mathcal{X}) = \sum_{t=1}^{T} f_t(x_t) - \min_{u \in \mathcal{X}} \sum_{t=1}^{T} f_t(u) \qquad (4)$$

$$\leq \sum_{t=1}^{T} \langle z_t, x_t - u_T \rangle, \qquad z_t \in \partial f_t(x_t), \quad (5)$$

where $u_T = \arg\min_{u \in \mathcal{X}} \sum_{t=1}^{T} f_t(u)$, and (5) provides an upper bound to the regret that is linear in $x_t$.

Assuming each $f_t$ is $L_t$-Lipschitz, $\mathcal{X} = \{x : ||x||_2 \leq B\}$, and $\eta = \frac{B}{L\sqrt{2T}}$, it has been shown that (5) implies that $regret_{OGD}^T \leq BL\sqrt{2T}$ where $L^2 \geq \frac{1}{T}\sum_{t=1}^{T} L_t^2$ (Shalev-Shwartz 2011).

## 3 Online Monotone Games & *Auto*-Welfare

In this section, we define monotone games and *auto*-welfare.

**Definition 1** (Monotone Game). *A game is monotone if the map, $F : \mathcal{X} \to \mathbb{R}^n$, formed by concatenating the subgradients of all $N$ player cost functions, $f^{(i)}(x)$, is monotone. More concretely, let $F = [z^{(1)}, \ldots, z^{(N)}]$ where $z^{(i)} \in \partial_{x^{(i)}} f^{(i)}$ is any subgradient of $f^{(i)}$ w.r.t. $x^{(i)}$. A game is monotone if $F$ satisfies (2).*

Essentially, a game is monotone if gradient descent with an infinitesimally small step size, e.g., GIGA (Zinkevich 2003), does not cause the player strategies to diverge away from the equilibrium point. In online monotone games, an adversary may choose a new monotone game for the players to play at every time step, i.e., $f^{(i)}(x)$ becomes $f_t^{(i)}(x)$.

We begin our discussion of regret with a trivial application of online optimization to games. Online optimization provides theory for bounding the regret of an algorithm's prediction, $x_t$, when comparing to a baseline, $u_T$, chosen in retrospect. In other words, if we were to go back in time and play this baseline against the same exact sequence of environments, how much better would we do? To measure this, we can sum up the differences between the algorithm's loss and the baseline loss at each time step, $t$, as in (4).

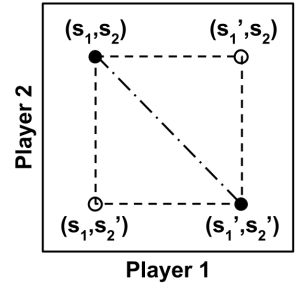

Figure 2: Illustrative comparison of *auto*-welfare to a *game-agnostic* loss. Online optimization provides theory for regret measured only along the edges of the square (axis aligned), while online monotone optimization additionally measures regret along diagonals (any line).

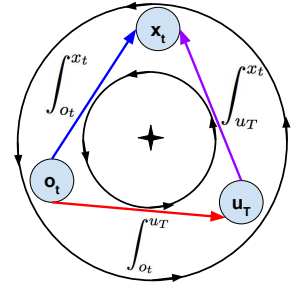

Figure 3: Illustrative comparison of *two-step*, $\int_{o_t}^{x_t} - \int_{o_t}^{u_T}$, to *one-step* regret, $\int_{u_T}^{x_t}$.

In order to use this theory in a game, we simply focus our attention on one player and treat the rest of the players as part of the adversarial environment. In this way, online optimization can provide regret bounds for each player if we imagine replaying the game but with all other players forced to replay the same actions as before. This is largely unsatisfying given that it seems to have taken the game aspect out of the game. Ideally, our regret measure would leave the game environment intact and allow all players to change their actions. In this regard, welfare regret is far more satisfying because it measures the sum of all player payoffs with respect to a baseline that allows all players to change their actions. Unfortunately, bounding welfare regret often requires properties like smoothness.

As a compromise, we propose *auto*-welfare. Consider player 1 in an $N$-player game. Player 1 receives a payoff or *reward* for changing her strategy, however, her reward depends on all other player adjustments as well. Player 1 never knows how the other $N-1$ players are going to change their strategies, so it is reasonable for her to measure the portion of her reward that is due to her strategy change alone. Such a measurement provides valuable feedback on her decision to update her strategy, and this measurement is exactly what *auto*-welfare sums for all players. Therefore, *auto*-welfare can be thought of as measuring how "satisfied" the players as a whole are with their decision making given that they only have control over themselves. In contrast, welfare measures how "satisfied" the players as a whole are with the outcome of the game.

We can compute *auto*-welfare, $W^a$, with a path integral,

$$W_t^a(x_t) = W_{o_t}^a + \int_{x:o_t \to x_t} \langle -F_t(x), dx \rangle, \quad (6)$$

where $F_t(x)$ is an output of the game map (see Definition 1), $o_t$ is any reference vector with known *auto*-welfare, $W_{o_t}^a$, and $x : o_t \to x_t$ is the straight line path from $o_t$ to $x_t$ through $\mathcal{X}$. Figure 2 illustrates the flexibility *auto*-welfare provides over the *game-agnostic* loss provided by online optimization theory. This formulation has been considered for converting symmetric VIs into equivalent optimization problems (Aghassi, Bertsimas, and Perakis 2006; Hu and Wang 2006), however, to our knowledge has not been leveraged for asymmetric VIs, which represent a wider class of games.

We can rewrite *auto*-welfare as follows to reveal its relationship to standard welfare, $W$.

$$W_t^a(x_t) = W_{o_t}^a + \int_{x:o_t \to x_t} \langle -F_t(x), dx \rangle \quad (7)$$

$$= W_{o_t}^a + \sum_{i=1}^N \int_{x:o_t \to x_t} \langle -\partial f_{t,i}^{(i)}(x), dx^{(i)} \rangle \quad (8)$$

$$= W(x_t) - \sum_{i=1}^N \sum_{j \neq i}^N \underbrace{\int_{x:o_t \to x_t} \langle -\partial f_{t,j}^{(i)}(x), dx^{(j)} \rangle}_{i\text{'s reward due to } j\text{'s strategy change}} \quad (9)$$

where $\partial f_{t,j}^{(i)}(x)$ is a subgradient of agent $i$'s expected loss function with respect to agent $j$'s strategy, $x_t^{(j)}$, evaluated at $x$. Therefore, $W_t^a(x_t)$ gives welfare minus the rewards resulting from intra-team inefficiencies. We call this *auto*-welfare because it sums the portions of the player's welfare that can be attributed to its own strategy.

### 3.1 Online Monotone Optimization

We are now ready to present a framework for Online Monotone Optimization (Algorithm 3: OMO) that will enable us to derive lower bounds for *auto*-welfare regret in online monotone games. Let $f_t(x_t) = -W_t^a(x_t)$ be the *path integral loss*.

---

**Algorithm 3** Online Monotone Optimization (OMO)

input: A convex set $\mathcal{X} \subseteq \mathbb{R}^n$
**for all** $t = 1, 2, \ldots$ **do**
  predict a vector $x_t \in \mathcal{X}$
  receive a monotone map, $F_t$, with
    reference vector, $o_t$, and reference scalar, $f_{o_t}$
  suffer $f_t(x_t) = -W_t^a(x_t)$ as defined in (6)
**end for**

---

Comparing OMO to OCO, we see that the major difference is that we now receive a loss function implicitly defined by a monotone map whereas in OCO, we receive a convex loss function directly. In some cases, OMO reduces to OCO, however, this is not always the case, so we cannot rely on OCO theory alone to bound *auto*-welfare. In general, OMO represents a strict superset of OCO (see Appendix A.4).

**Theorem 1.** *$OCO(f_t, \mathcal{X})$ is equivalent to $OMO(\partial f_t, \mathcal{X})$ and $\exists F_t$ such that $OMO(F_t, \mathcal{X}) \notin \{\forall f_t \ OCO(f_t, \mathcal{X})\}$ implying $OCO \subset OMO$ in the strict sense.*

However, if $F$ is affine, OMO is equivalent to OCO (see Appendix A.5).

**Theorem 2.** *If $F_t(x_t) = Ax_t + b$ and $A$ is PSD, then $OMO(F_t, \mathcal{X})$ is equivalent to $OCO(f_t, \mathcal{X})$.*

Next, we revisit the definition of regret given by (4). Two definitions of regret naturally arise in the OMO framework. Both notions are equivalent to (4) in the OCO setting (i.e., 1-player monotone games). In the following discussion, we refer to $\int_{x:a \to b} \langle F_t(x), dx \rangle$ with the shorthand $\int_a^b$; note that this measures how much worse strategy set $b$ is relative to $a$. Please consult Figure 3 for a visual during this discussion.

*Two-step* regret compares the loss of the online learning strategy to the loss of the best fixed strategy in retrospect ($\int_{o_t}^{x_t} - \int_{o_t}^{u_T}$). *One-step* regret represents the loss felt when converting the best fixed strategy to the online learning strategy ($\int_{u_T}^{x_t}$). While these two notions are equivalent in OCO, they are possibly different in OMO due to path-dependence of general monotone maps. We opt for *one-step* regret as it lacks any direct dependence on $o_t$. First we provide a useful integral upper bound over monotone maps (see Remark 3.10 in the work of Romano et al. (1993) for a formal proof).

**Lemma 1** (Path Integral Bound). *The path integral over a monotone map is bounded by its linear approximations, i.e., $\langle F(a), b - a \rangle \leq \int_{x:a \to b} \langle F, dx \rangle \leq \langle F(b), b - a \rangle.$*

Using this, we can immediately derive linear upper bounds for *two-step* and *one-step* regret as follows:

$$\text{regret}_2 = \int_{o_t}^{x_t} - \int_{o_t}^{u_T}$$
$$\leq \langle F(x_t), x_t - o_t \rangle - \langle F(o_t), u_T - o_t \rangle, \quad (10)$$

$$\text{regret}_1 = \int_{u_T}^{x_t} \leq \langle F(x_t), x_t - u_T \rangle$$
$$\leq \langle F(x_t), x_t - o_t \rangle - \langle F(x_t), u_T - o_t \rangle. \quad (11)$$

Unfortunately, we cannot simplify the difference between the latter terms, $\langle F(x_t) - F(o_t), u_T - o_t \rangle$, and so it is not clear if one of these regrets upper bounds the other. However, we can bound the difference between the two regrets using Stokes' theorem and bounds on $F$ and its derivatives. The difference between the two regrets is equal to the magnitude of the path integral around the triangle in Figure 3. By Stokes' theorem,

$$\left| \oint_{\partial \Sigma} \langle F, dx \rangle \right| = \left| \int_\Sigma \nabla \times F \cdot d\Sigma \right| \leq \max_\Sigma ||\nabla \times F|| \cdot \int_\Sigma dA$$
$$\leq 2\sqrt{2(\beta^2 + L\gamma)} \cdot (\text{Area of } \triangle) \quad (12)$$

where $L$, $\beta$, and $\gamma$ are bounds on $F$, the Jacobian of $F$, and a matrix of 2nd derivates respectively and $\Sigma$ ($\partial \Sigma$) is the 2-dimensional area (perimeter) formed by the path around the triangle, $\triangle$ (see Appendix A.8 for proof).

We summarize the new definitions and bounds using regret$_1$ here for convenience.

$$regret_{\mathcal{A}}^{(t,T)}(\mathcal{X}) = \int_{x:u_T \to x_t} \langle F_t(x), dx \rangle$$
$$\leq \langle F_t(x_t), x_t \rangle - \langle F_t(x_t), u_T \rangle \quad (13)$$

$$regret_{\mathcal{A}}^T(\mathcal{X}) = \sum_{t=1}^T regret_{\mathcal{A}}^{(t,T)}(\mathcal{X}) \quad (14)$$

$$u_T = \arg\min_{u \in \mathcal{X}} \sum_{t=1}^T \int_{x:o_t \to u} \langle F_t(x), dx \rangle \quad (15)$$

### 3.2 Derivation of No-Regret Algorithms for OMO

Due to the simplicity of the instantaneous regret bounds discussed in the previous section and the work previously done in OCO, the derivation of no-regret algorithms for OMO is trivial. We have shown that instantaneous regret for general monotone maps can be bounded above by considering the constant approximation of the map in (13). Note that a constant map, $F_t(x_t)$, is always the subgradient of some linear function, $f_t(x) = \langle F_t(x_t), x \rangle$. This implies that the regret for general monotone maps is bounded above by considering the online linear optimization problem with $f_t(x)$. This reduction mirrors that of OCO, where convex losses are bounded above by their linear approximations. The implication is that the online gradient decent and even online mirror descent algorithms can be adapted from OCO with almost no effort to minimize regret in OMO while enjoying exactly the same $o(T)$ regret bounds (see end of Section 2 and Appendix A.7). This is somewhat surprising as Theorem 1 indicates that OMO sometimes involves minimizing non-convex functions. The no-regret algorithms for OMO are given in Algorithms 4 and 5.

Note that OMO defines $f_t(x_t) = -W_t^a(x_t)$. No-regret algorithms for OMO minimize an upper bound on regret as defined in (14) which immediately translates into a lower bound for *auto*-welfare regret. Therefore, Algorithms 4 and 5 equivalently maximize a lower bound for *auto*-welfare.

Assuming each $F_t$ is $L_t$-Lipschitz, $\mathcal{X} = \{x : ||x||_2 \leq B\}$, and $\eta = \frac{B}{L\sqrt{2T}}$, it directly follows from (Shalev-Shwartz 2011) that $regret_{OMoMD}^T(\mathcal{X}) \leq BL\sqrt{2T}$ where $L^2 \geq \frac{1}{T}\sum_{t=1}^T L_t^2$. Moreover, if $o_t = x_{t-1}$, then the difference between regret$_1$ and regret$_2$ is bounded by (12), which decays at $\mathcal{O}(T^{-1/2})$. In this case, both average regret$_2$ and average regret$_1$ go to zero in the limit, $T \to \infty$, i.e., (Area of $\triangle$) $\leq \frac{1}{2}||x_t - x_{t-1}|| \cdot ||x_t - u_T|| \leq \frac{1}{2}(\eta L)(2B) = \frac{B^2}{\sqrt{2T}}$.

### 3.3 Summary

A monotone game is a game where the map formed by concatenating all player gradients is a monotone map. *Auto*-welfare measures the portion of the players' welfare that can be attributed to their own strategy and provides a sense of how "satisfied" the players are with their decision making. Moreover, *auto*-welfare can be computed as a path integral over a monotone game map and is lower bounded by a linear approximation (see Lemma 1). The Online Monotone

---

**Algorithm 4** Online Monotone Descent (OMoD)
  input: A scalar learning rate $\eta > 0$
  $x_1 = 0$
  **for all** $t = 1, 2, \ldots$ **do**
    $x_{t+1} = x_t - \eta z_t$ where $z_t \in F_t(x_t)$
  **end for**

---

**Algorithm 5** OMoMirrorD (OMoMD)
  input: A link function $g : \mathbb{R}^n \to \mathcal{X}$
  $x_1 = g(0)$
  **for all** $t = 1, 2, \ldots$ **do**
    $\theta_{t+1} = \theta_t - \eta z_t$ where $z_t \in F_t(x_t)$
    $x_t = g(\theta_{t+1})$
  **end for**

---

Optimization framework formalizes the problem of maximizing *auto*-welfare in an environment where the game itself may change at each step. In a "changing game", player losses/utilities may be arbitrarily changed as long as the resulting game is still monotone. This is equivalent to arbitrarily removing, substituting, and/or adding players at each step. By leveraging OCO theory, we show that natural gradient descent dynamics (OMoMD) ensure sublinear average *auto*-welfare regret even in this extreme scenario.

## 4 AGT Background and Related Work

Monotone games and socially-convex games are both subsets of convex games (see Theorems 7 and 8 in Appendix A.9). Convex games are games in which each agent's cost function, $f^{(i)}(x)$, is convex with respect to its own strategy, $x^{(i)}$. Gordon, Greenwald, and Marks (2008) studied internal and external regret for individual agents in convex games and related these to convergence towards correlated and coarse-correlated equilibria respectively. Note that in general, these results on equilibria do not imply results for welfare. Even-Dar, Mansour, and Nadav (2009) examined no-regret algorithms in *socially-concave games* (equivalently formulated as *socially-convex* games), and showed that each player's average strategy approaches that player's strategy at the Nash equilibrium; also, each player's average utility approaches that player's utility at the Nash equilibrium. Roughgarden (2009) developed the notion of *smooth* games. Smoothness relates the convergence of strategies towards Nash equilibria to the *price of anarchy* (PoA), which defines the ratio of the worst-case sum cost of a Nash equilibrium, $\max_{x^* \in \mathcal{X}^*} -W(x^*)$, to the best-case sum cost of a player strategy set, $\min_{x \in \mathcal{X}} -W(x)$. In short, smoothness relates no-regret dynamics to the welfare of a game. Table 1 outlines the intersections between the various game types.

Additional performance gains can be obtained if we can assume each player in the game is employing an algorithm from a given class. Syrgkanis et al. (2015) have accelerated convergence to Nash equilibria, to zero-regret, and to optimal welfare (assuming the game is smooth) under this scenario. An extension of this work also considers scenarios where the game is changing at each time step (Lykouris, Syrgkanis, and Tardos 2016). Specifically, Foster et al. (2016) showed that approximate optimality can be guaranteed if the game allows players to be replaced with probability $p$ for small $p$. Moreover, this is true even when the players observe only bandit feedback (as opposed to expected costs) when comparing against a dynamic baseline. Critically, bounds on welfare are still derived from a smoothness constraint on the game.

| Type / Ex. | a | b | c | d | e | f | g | h | i |
|---|---|---|---|---|---|---|---|---|---|
| Smooth | √ | √ | √ | √ | √ | | | | |
| Convex | | √ | √ | √ | √ | √ | √ | √ | √ |
| Monotone | | | √ | | √ | | √ | | √ |
| Socially-Convex | | | | √ | √ | | | √ | √ |

Table 1: Games may share multiple properties at once. Definitions of properties and examples for each case (denoted by the column heading) are given in Appendix A.9.

As a whole, the game properties discussed above (convex, socially-convex, smooth), focus on player costs. In contrast, we examine *monotone* games where monotonicity is a property of game dynamics—in other words, a property of the player algorithms that are employed. However, in this paper, when we refer to a game as monotone, we are stating that the game dynamics associated with all players running gradient descent with the same learning rate parameter are monotone. We defer progress into understanding more general monotone algorithms to future research.

## 5 Applications

We illustrate applications of online monotone game theory on modeling several concave games, solving an online variational inequality problem (OVI), uncovering insights into a saddle-point based reinforcement learning algorithm, and designing a generative adversarial network (GAN), a game-based generative model, with *auto*-welfare regret guarantees.

### 5.1 Concave Games

Even-Dar, Mansour, and Nadav (2009) developed the theory of socially-concave games and showed that minor simplifications of a number of concave games are socially-concave as well. In Appendix A.10, we show that variants of these games also satisfy monotonicity. Specifically, we prove $F$ is monotone for the following games:

- Linear *Cournot* Competition (Cournot 1838)—Firms compete for consumers by adjusting quantities of goods produced. The price of goods is set by a linear function of the total quantity of goods in the market.

- Linear Resource Allocation (Johari and Tsitsiklis 2004)—A network controller oversees the sharing of a communication channel, ensuring total communication does not exceed the network capacity. Users submit bids to the controller, which the controller uses when deciding how to allocate capacity. User value functions are linear.

- Congestion Control Protocols (Even-Dar, Mansour, and Nadav 2009)—We consider a *Tail Drop* policy where a router drops packets that exceed the network capacity.

In Appendix A.10, we analyze resource allocation and compare welfare with *auto*-welfare. In this case, welfare is maximized when all users submit the minimum bid amount. This result is independent of the parameters of the users' utility functions, which from a modeling viewpoint is unsatisfying. In contrast, *auto*-welfare is maximized by bids with an intuitive dependence on the utility function parameters: 1) as the penalty for large bids grows, the optimal bid amount decreases; 2) as the number of users increases, the optimal bid amount increases (due to increased competition), approaching an asymptote.

### 5.2 A Machine Learning Economy

Next, we consider a cloud-based machine learning network (MLN) adopted from (Nagurney and Wolf 2014). Providers of machine learning data control the quantity of data provided while network providers control the delivery price as well as service quality. Consumers influence the network through demand functions dictating the prices they are willing to pay for specific quantities and qualities of services rendered. See Appendix A.11 for a more thorough description.

We define each firm's utility function to be concave and quadratic in its strategy. This establishes the equivalence between the equilibrium state we are searching for and the variational inequality to be solved, $VI(F_t, \mathbb{R}^+)$, where $F_t$ returns a vector consisting of the negative gradients of the utility functions for each firm.

To cast this VI as an online learning problem, we let the parameters of the network change. This creates a more realistic model as a number of external factors can cause the network to change such as complex network congestion effects, cyber attacks, etc. The goal then is to predict the equilibrium point of each new MLN in the face of these possibly adversarial forces. Specifically, our experiment considers ten different five-firm networks with monotone $F_t$. At each time step, the adversary receives OMoD's prediction for the equilibrium point and returns the MLN whose equilibrium is farthest from the predicted one. The reference vectors, $o_t$, are the previous strategy sets, $x_{t-1}$. The reference constants, $f_t(o_t)$, are all assumed to be zero without loss of generality.

Computing the optimal strategy $u_T$ is prohibitive in general (15). We cannot even say that $u_T$ lies in the convex hull of the individual equilibria, $x_t^*$, due to $f_t = -W_t^a$ being possibly non-convex. Instead, we approximate $u_T$ with the equilibrium of the network created from averaging the ten together, $\bar{x}^*$. We measure two regrets throughout the learning process. The first is an approximation of *auto*-welfare regret using regret$_1$, and the second uses regret$_2$. Figure 4 plots the average of these measures with respect to the time step, demonstrating that both average *auto*-welfare regrets approach zero in support of our derived sublinear bounds.

### 5.3 GTD Algorithms

Reinforcement Learning (RL) is a class of learning problems in which an agent attempts to maximize a long-term reward in an unfamiliar environment by reinforcing rewarding behaviors. Solving this problem typically requires first learning a *value function*, $V^\pi(s)$, which gives the long-term reward the agent is expected to receive if employing policy $\pi$ and starting from state $s$. Often, we learn an approximate value function instead, $V^\pi_\theta(s)$, parameterized by $\theta \in \mathbb{R}^d$. Approximating $V^\pi_\theta(s)$ by observing an exploratory policy, $\pi'$, is called *off-policy policy evaluation*. The gradient temporal difference (GTD) algorithms form a family of algorithms for accomplishing this task (Sutton, Maei, and Szepesvári 2009; Sutton et al. 2009).

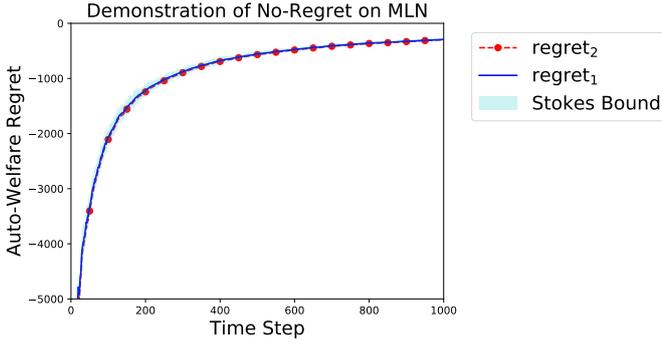

Figure 4: Demonstration of OMoMD on described machine learning network. Note that regret$_1$ and regret$_2$ fall within the light blue region predicted by the Stokes bound in (12).

Liu et al. (2015) showed that although the family of gradient temporal difference (GTD) algorithms are technically not gradient algorithms with respect to their original objectives, they are gradient algorithms with respect to a saddle point objective, which is an example of a 2-player game. We show that this game is strongly monotone (see Appendix A.12). The GTD update rules are given by

$$y_{t+1} = y_t + \alpha_t(b - A\theta_t - My_t) \quad (16)$$
$$\theta_{t+1} = \theta_t + \alpha_t(A^T y_t) \quad (17)$$

where $M = \mathbb{1}_d$ or $M$ is a covariance matrix. These updates are equivalent to running OMoD on an appropriate 2-player game. Either way, $M$ is symmetric positive definite and the corresponding map, $F$, is strongly monotone with parameter 1 for $M = \mathbb{1}_d$, or more generally, $\lambda_{\min}(M)$.

The corresponding path integral loss that GTD bounds is

$$f([y;\theta]) = \frac{1}{2}\underbrace{(y^T My - y_0^T My_0)}_{y\text{-growth}} - \underbrace{y^T(b - A\theta_0) + y_0^T(b - A\theta)}_{y\text{-alignment}} \quad (18)$$

In the GTD algorithms, $y$ was originally introduced as an auxiliary variable to estimate $\mathbb{E}[\rho\delta\phi] = b - A\theta$. If we let $\theta_0 = \theta$, and both $y$-growth$<0$ and $y$-alignment$>0$, this means that $y$ is a better estimate of $b - A\theta$ than $y_0$. This is reassuring because it means that the GTD algorithms are ensuring a bound on the error of this estimate. Accurately estimating $b - A\theta$ was key to deriving the GTD algorithms, however, this was motivated out of intuition. We hope this perspective provides a stronger mathematical explanation as well as bounds on the estimation error.

### 5.4 Affine Wasserstein GANs

Generative adversarial networks formulate the training of a generative model as a game (Goodfellow et al. 2014). The original formulation is a minimax game between a generator, $G(z) : z \to x$, and a discriminator, $D(x) : x \to [0, 1]$,

$$\min_G \max_D V(D, G) = \mathbb{E}_{x \sim p_{data}(x)}\Big[\log(D(x))\Big]$$
$$+ \mathbb{E}_{z \sim p_z(z)}\Big[\log(1 - D(G(z)))\Big], \quad (19)$$

where $p_{data}(x)$ is the true data distribution and $p_z(z)$ is a simple (usually fixed) distribution that is easy to draw samples from (e.g., $\mathcal{N}(0, 1)$).

Unfortunately, this game is not monotone in general, however, we can derive a monotone version of the Wasserstein-type GAN (Arjovsky, Chintala, and Bottou 2017) with online guarantees using regret$_1$. The new minimax objective is

$$\min_G \max_d V(G, d) = \mathbb{E}_{x \sim p_{\text{data}}(x)}[d^T x] - \mathbb{E}_{z \sim p_z(z)}[d^T Gz] \quad (20)$$

where $x \in R^n, z \in R^m, d \in R^n, G \in R^{n \times m}$. We derive the map, $F$, associated with this game in Appendix A.13 (299). $F$ is monotone for any $p_z(z)$ and $p_{\text{data}}(x)$. If $G$ and $d$ are regularized with $\frac{\alpha}{2}||\cdot||_2^2$, then $F$ is strongly monotone with parameter $\alpha$.

In Appendix A.14, we show that for any differentiable, convex-concave minimax game, the corresponding path integral loss that Algorithms 4 and 5 bound is

$$f([G; d]) = V(G, d_0) - V(G_0, d). \quad (21)$$

If we let $d_0 = d$, then this loss is minimized if $G$ tricks $d$ better than $G_0$ tricks $d$; i.e., the generator, $G$, is an improvement over $G_0$. Alternatively, if we let $G_0 = G$, then this loss models how much the discriminator, $d$, has improved over $d_0$. Furthermore, if $(d_0, G_0) = (d^*, G^*)$, i.e. the minimax solution, then $f$ is a familiar Lyapunov function for the game.

## 6 Conclusion & Future Work

We proposed a new framework for online learning, namely Online Monotone Optimization, which enables the study of *auto*-welfare regret for online monotone games. This framework generalizes the popular Online Convex Optimization framework in a way that allows it to model regret for multiple agents in parallel while still retaining the simplicity of standard no-regret algorithms from previous work. We support the efficacy of our new framework with connections to network congestion protocols, empirical results from a VI, analysis of an existing RL algorithm, and design of a generative model.

In future work, we hope to develop the framework further to handle dynamic regret. We also hope to develop efficient algorithms for approximating the optimal strategy $u_T$ in retrospect.

# Contents

# A Appendix

## A.1 Monotonicity in Integral Form

A proof sketch for Lemma 1.

*Proof.* Let $x_{i+1} - x_i = \frac{x_n - x_0}{n} \ \forall \ x_0, x_n$ and recall the definition of monotonicity, $\langle F(x_{i+1}) - F(x_i), x_{i+1} - x_i \rangle \geq 0 \ \forall \ x_i, x_{i+1}$ which implies

$$\langle F(x_i), \frac{x_n - x_0}{n} \rangle \leq \langle F(x_{i+1}), \frac{x_n - x_0}{n} \rangle \tag{22}$$

which implies

$$\langle F(x_i), \frac{x_n - x_0}{n} \rangle \leq \langle F(x_j), \frac{x_n - x_0}{n} \rangle \ \forall \ j \geq i. \tag{23}$$

Also,

$$\langle F(x_0), x_n - x_0 \rangle = \langle F(x_0), \sum_{i=0}^{n-1} \frac{x_n - x_0}{n} \rangle \tag{24}$$

$$= \sum_{i=0}^{n-1} \langle F(x_0), \frac{x_n - x_0}{n} \rangle \tag{25}$$

$$\leq \sum_{i=0}^{n-1} \langle F(x_i), \frac{x_n - x_0}{n} \rangle \tag{26}$$

$$= \int_{x:x_0 \to x_n} \langle F, dx \rangle \text{ as } n \to \infty, \tag{27}$$

and vice versa for the reverse direction, which implies

$$\implies \langle F(x_0), x_n - x_0 \rangle \leq \int_{x:x_0 \to x_n} \langle F, dx \rangle \tag{28}$$

$$\leq \langle F(x_n), x_n - x_0 \rangle. \quad \square \tag{29}$$

$\square$

## A.2 Pseudo-monotonicity in Integral Form

**Definition 2** (Pseudo-monotone). *$F$ is pseudo-monotone if the following one-way implication holds for all $x, y \in K$: $\langle F(x), y - x \rangle \geq 0 \implies \langle F(y), y - x \rangle \geq 0$.*

**Lemma 2.** *If $F$ is pseudo-monotone, $F$ also obeys the following one-way implication*

$$\langle F(x), y - x \rangle \geq 0 \implies \int_{z:x \to y} \langle F(z), dz \rangle \geq 0 \tag{30}$$

*Proof.* Assume $\langle F(x), y - x \rangle \geq 0$ and let $\Delta z = \frac{y-x}{n}$. Then

$$\langle F(x), y - x \rangle = \langle F(x), \frac{y-x}{n} \rangle \cdot n \tag{31}$$

$$= \langle F(x), \Delta z \rangle \cdot n \tag{32}$$

$$\geq 0 \tag{33}$$

$$\implies \langle F(x), i\Delta z \rangle \geq 0 \ \forall \ i \geq 0 \tag{34}$$

Let $i\Delta z = (x + i\Delta z) - x = \hat{y}_i - x$.

$$\langle F(x), i\Delta z \rangle \geq 0 \ \forall \ i \geq 0 \tag{35}$$

$$\implies \langle F(x), \hat{y}_i - x \rangle \geq 0 \ \forall \ i \geq 0 \tag{36}$$

$$\implies \langle F(\hat{y}_i), \hat{y}_i - x \rangle \geq 0 \ \forall \ i \geq 0 \tag{37}$$

$$\implies \langle F(\hat{y}_i), i\Delta z \rangle \geq 0 \ \forall \ i \geq 0 \tag{38}$$

$$\implies \langle F(\hat{y}_i), \Delta z \rangle \geq 0 \ \forall \ i \geq 0 \tag{39}$$

$$\implies \sum_{i=0}^{n} \langle F(\hat{y}_i), \Delta z \rangle \geq 0 \tag{40}$$

$$\text{where } \Delta z = \Delta z(n) \tag{41}$$

$$\implies \lim_{n \to \infty} \sum_{i=0}^{n} \langle F(\hat{y}_i), \Delta z \rangle \geq 0 \tag{42}$$

$$= \int_{z:x \to y} \langle F(z), dz \rangle \geq 0 \tag{43}$$

$\square$

## A.3 Nash Equilibrium vs VI Solution

**Theorem 3.** *Repeated from (Cavazzuti, Pappalardo, and Passacantando 2002). Let $(\mathbf{C}, K)$ be a cost minimization game with player cost functions $C_i$ and feasible set $K$. Let $x^*$ be a Nash equilibrium. Let $F = [\frac{\partial C_1}{\partial x_1}, \ldots, \frac{\partial C_N}{\partial x_N}]$. Then*

$$\langle F(x^*), x - x^* \rangle \geq 0 \tag{44}$$
$$\forall x \in (\{x^* + \mathbf{I}_K(x^*)\} \cap K) \subseteq K \tag{45}$$

where $\mathbf{I}_K(x^*)$ is the internal cone at $x^*$. When $C_i(\mathbf{x}_i, \mathbf{x}_{-i})$ is pseudo-convex in $\mathbf{x}_i$ for all $i$, this condition is also sufficient.

## A.4 Theorem 1: OCO ⊂ OMO

Let the feasible set, $\mathcal{X}$, and field, $F(x)$, be defined as follows.

$$x = [r, c] \in \mathcal{X} \equiv [0, 1]^2 \tag{46}$$
$$F(x) = \begin{pmatrix} r^2 + 2rc + c^2 \\ -2r^2 + 2rc + c^2 \end{pmatrix} \tag{47}$$

In addition, we set the reference vector, $o$, to the origin, $[0, 0]$. We set the reference scalar, $f(o)$, to 0 without loss of generality.

**$F$ is monotone over $\mathcal{X} = [0, 1]^2$**

$$J(F) = \begin{pmatrix} 2r+2c & 2r+2c \\ -4r+2c & 2r+2c \end{pmatrix} \tag{48}$$

$$J_s(F) = \frac{1}{2}(J + J^T) = \begin{pmatrix} 2r+2c & 2c-r \\ 2c-r & 2r+2c \end{pmatrix} \tag{49}$$

$$det(J_s) = 3r^2 + 12rc \geq 0 \;\forall\; [r, c] \in \mathcal{X} \tag{50}$$
$$\tau(J_s) = 4r + 4c \geq 0 \;\forall\; [r, c] \in \mathcal{X} \tag{51}$$
$$\implies J_s(F) \succeq 0. \tag{52}$$

The trace and determinant of the $(2 \times 2$ matrix) symmetrized $J_s$ are both non-negative, which imply the eigenvalues of $J_s$ are non-negative. Therefore, $F$ is monotone. □

**$f$ is non-convex over $\mathcal{X} = [0, 1]^2$**

$$f(x) = \cancel{f_o} + \int_{z:o \to x} \langle F, dz \rangle \tag{53}$$

$$= \int_0^1 \langle F(o + \tau(x - o)), (x - o) d\tau \rangle \tag{54}$$

$$= \int_0^1 \langle F(\tau x), x \rangle d\tau \tag{55}$$

$$= \frac{1}{3}(r^3 + 3rc^2 + c^3) \tag{56}$$

$$H(f) = \begin{pmatrix} 2r & 2c \\ 2c & 2r+2c \end{pmatrix} \tag{57}$$

$$det(H) = 4(r^2 + rc - c^2) < 0 \tag{58}$$

$$\forall \; \{[r, c] \,|\, [r, c] \in \mathcal{X}, c > \frac{\sqrt{5}+1}{2}r\} \tag{59}$$

$$\implies H \not\succeq 0. \tag{60}$$

This means $f$ forms a saddle surface over a compact subset of $\mathcal{X}$, therefore, it is non-convex. In fact, $f$ is not even quasi-convex. For example, let $x_0 = [0, 0.8], x_f = [0.5, 0.45]$ and consider their midpoint, then

$$f(\frac{x_0 + x_f}{2}) \not\leq \max\{f(x_0), f(x_f)\}. \tag{61}$$

## A.5 Theorem 2: OMO ≡ OCO for Positive semidefinite Affine Maps

This concerns such problems as linear complementarity problems (LCPs).

$$F_t(x_t) = Ax_t + b, \tag{62}$$
$$\text{where } x_t, b, o_t \in \mathbb{R}^n, \; A \succeq 0 \in \mathbb{R}^{n \times n} \tag{63}$$

$$f_t(x_t) - f_{o_t} = \tag{64}$$

$$= \int_{x:o_t \to x_t} \langle F_t, dx \rangle \tag{65}$$

$$= \int_0^1 \langle F_t(o_t + \tau(x_t - o_t)), (x_t - o_t) d\tau \rangle \tag{66}$$

$$= \int_0^1 \langle A(o_t + \tau(x_t - o_t)) + b, (x_t - o_t) d\tau \rangle \tag{67}$$

$$= \int_0^1 \langle Ao_t + \tau A(x_t - o_t) + b, (x_t - o_t) d\tau \rangle \tag{68}$$

$$= \int_0^1 \langle Ao_t + b, (x_t - o_t) d\tau \rangle + \tau \langle A(x_t - o_t), (x_t - o_t) d\tau \rangle \tag{69}$$

$$= \langle Ao_t + b, (x_t - o_t) \rangle + \frac{1}{2} \langle A(x_t - o_t)), (x_t - o_t) \rangle \tag{70}$$

$$= o_t^T A^T x_t - o_t^T A^T o_t + b^T(x_t - o_t) + \frac{1}{2}(x_t - o_t)^T A^T (x_t - o_t) \tag{71}$$

$$= o_t^T A^T x_t - o_t^T A^T o_t + b^T(x_t - o_t) \frac{1}{2}[x_t^T A^T x_t - o_t^T A^T x_t - x_t^T A^T o_t + o_t^T A^T o_t] \tag{72}$$

$$= \frac{1}{2}[x_t^T A^T x_t + x_t^T(A - A^T)o_t - o_t^T A^T o_t] + b^T(x_t - o_t) \tag{73}$$

$$= \frac{1}{2}[x_t^T \left(\frac{A + A^T}{2}\right) x_t + x_t^T(A - A^T)o_t - o_t^T A^T o_t] + b^T(x_t - o_t) \tag{74}$$

$$\tag{75}$$

$$Hessian(f_t) = \frac{1}{2}[A + A^T] \succeq 0 \implies f_t \text{ is convex} \tag{76}$$

This also implies that every multivariate function with nonzero Hessian can be represented by an infinite number of fields (other than the gradient), specifically any field whose symmetric component equals $\frac{A+A^T}{2}$.

## A.6 Monotone Optimization with o = x*

Here, we consider the case where the reference point is the solution to the corresponding variational inequality problem, $o = x^* = VI(F, \mathcal{X})$. Remember, this means $x^*$ is an equilibrium point of the field, $F$, and has the property

$$\langle F(x^*), z - x^* \rangle \geq 0 \ \forall z \in \mathcal{X}. \tag{77}$$

**MoMD solves MO when** $o = x^*$

**Theorem 4.** *If o is a solution to $VI(F, \mathcal{X})$ where $F : \mathcal{X} \to \mathbb{R}^n$ is a monotone (or at least pseudo-monotone) map and $\mathcal{X}$ is a convex set, then o is a global minimizer of the monotone optimization problem with map $F$, reference vector o, and any reference scalar $f_{o_t}$.*

*Proof.* Without loss of generality, let $f(o) = 0$.

$$\nabla_x \Big\{ \int_{z:o \to x} \langle F(z), dz \rangle \Big\} \Big|_{x=o} = \tag{78}$$

$$= \nabla_x \Big\{ \int_{t:0 \to 1} \langle F(o + t(x-o)), x - o \rangle dt \Big\} \Big|_{x=o} \tag{79}$$

$$= \int_{t:0 \to 1} \Big\{ F(o + t(x-o)) + \tag{80}$$

$$J(o + t(x-o))^T (x-o) t dt \Big\} \Big|_{x=o} \tag{81}$$

$$= \int_{t:0 \to 1} \Big\{ F(o) \Big\} \tag{82}$$

$$= F(o) \tag{83}$$

A necessary first order condition for optimality is

$$\langle F(o), z - o \rangle \geq 0 \; \forall z \in \mathcal{X}, \tag{84}$$

which by the definition of the variational inequality problem is solved by $o = x^*$. Reversing this result

$$f(o) = 0 \leq \langle F(o), z - o \rangle \leq \int_{x:o \to z} \langle F(x), dx \rangle = f(x) \tag{85}$$

reveals that $o = x^*$ is also a global minimum. $\square$

This directly implies that the Projection Method (MoMD with $R = \frac{1}{2}||x^2||$) converges exponentially fast to a minimum of the MO loss for strongly monotone fields.

Note that the optimality proof also carries through for pseudo-monotone fields (see A.2).

$$f(o) = 0 \leq \langle F(o), z - o \rangle \tag{86}$$

$$\implies 0 \leq \int_{x:o \to z} \langle F(x), dx \rangle = f(x) \tag{87}$$

which reveals that $o = x^*$ is also a global minimum of the pseudo-monotone loss.

### A.7 OMoD & OMoMD Regret Bounds

We repeat the bounds adopted from (Shalev-Shwartz 2011) for convenience.

**Theorem 5.** *Let $R$ be a $(1/\eta)$-strongly-convex function over $\mathcal{X}$ with respect to a norm $||\cdot||$. Assume that OMoMD is run on the sequence of monotone maps, $F_t$, with the link function*

$$g(\theta) = \arg\max_{x \in \mathcal{X}}(\langle x, \theta \rangle - R(x)). \tag{88}$$

*Then, for all $u \in \mathcal{X}$,*

$$\text{regret}_{OMoMD_T}(\mathcal{X}) \leq R(u_T) - \min_{v \in \mathcal{X}} R(v) + \eta \sum_{t=1}^T ||z_t||_*^2 \tag{89}$$

*Furthermore, if $F_t$ is $L_t$-Lipschitz with respect to $||\cdot||$, then we can further upper bound $||z_t||_* \leq L_t$.*

*Proof.* As we have shown previously,

$$\text{regret}_{A_{(t,T)}}(\mathcal{X}) = \int_{x:u_T \to x_t} \langle F_t, dx \rangle \tag{90}$$

$$\leq \langle F_t(x_t), x_t \rangle - \langle F_t(x_t), u_T \rangle \tag{91}$$

and the OMoMD algorithm is equivalent to running Follow the Regularized Leader (FoReL) on the sequence of linear functions with the regularization $R(x)$. The theorem now follows directly from Theorem 2.11 and Lemma 2.6 in (Shalev-Shwartz 2011). $\square$

OMoD is equivalent to OMoMD with $R(x) = \frac{1}{2}||x||_2^2$ and so the proof for OMoMD extends to OMoD as well.

## A.8 Curl Bound

The proof outline is as follows. The triangle in Figure 3 defines a 2-d plane in an $n$-dimensional ambient space. We are interested in the path integral around the triangle. Each element of the path integral consists of an inner product of the field $F$ with a differential vector along the curve. Any components of $F$ that are orthogonal to this differential vector evaluate to zero. The triangle is 2-d hence, its perimeter is 2-d, which means we may consider a projection of $F$, $F_{:2}$, onto the 2-d plane defined by the triangle. We can think of this projection as a rotation of $F$, $F^R = R \cdot F$, followed by a projection in which we extract the first 2-dimensions of $F_{:2} = \Pi_\Delta(F^R)$. The curl is defined for 3-d so we will actually append a third dimension whose component is identically zero. Define $F'$ to be this augmented projection of the field. Also, let $J$ be the Jacobian of a vector field $F$ (an $n \times n$ square matrix) and $J^2$ be a matrix of the same size but with second derivatives. Concretely,

$$J_{ij}(F) = \frac{\partial F_i}{\partial x_j} \qquad J^2_{ij}(F) = \frac{\partial^2 F_i}{\partial x_j^2}. \tag{92}$$

Define $J^{(2)}$ to be the 2nd principal submatrix of $J$, meaning the top left 2 block submatrix of $J$. Also, we define the following matrix norms

$$||A||_F = \Big(\sum_i \sum_j |A_{ij}|^2\Big)^{1/2} \tag{93}$$

$$||A||_2 = \sqrt{\lambda_{\max}(A^*A)} = \sigma_{\max}(A) \tag{94}$$

$$||A||_{2,1} = \sum_j \Big(\sum_i |A_{ij}|^2\Big)^{1/2} = \sum_j ||A_{\cdot j}||_2 \tag{95}$$

$$||A||_{\max} = \max_{ij} |A_{ij}| \tag{96}$$

which obey the following inequalities

$$||A||_{\max} \leq ||A||_2 \leq ||A||_F \leq n||A||_{\max}. \tag{97}$$

The following Lemmas are useful for the proof. We also define the following scalar values: $L, \beta, \gamma$.

**Lemma 3.** $||F'||_2 = ||F_{:2}||_2 \leq ||F^R||_2 \leq ||F||_2 = L$.

*Proof.*

$$\begin{aligned}
||F'||_2 &= ||F_{:2}||_2 & \text{because } F' \text{ is simply } F_{:2} \text{ augmented with a zero} & \quad (98) \\
&\leq ||F^R||_2 & \text{because the projection } \Pi_\Delta \text{ simply drops entries} & \quad (99) \\
&= ||RF||_2 & \text{by definition} & \quad (100) \\
&\leq ||R||_2 ||F||_2 & \text{because the } L_2 \text{ matrix norm is submultiplicative} & \quad (101) \\
&= ||F||_2 & \text{because the spectral norm of rotation matrix is 1} & \quad (102)
\end{aligned}$$

$\square$

**Lemma 4.** $||J(F')||_F^2 \leq 4||J(F)||_2^2 = 4\beta^2$.

*Proof.*

$$\begin{aligned}
||J(F')||_F^2 &= ||J(F_{:2})||_F^2 & \text{because } J(F') \text{ is simply } J(F_{:2}) \text{ augmented with zeros} & \quad (103) \\
& & \text{along the third row and third column.} & \\
&\leq 4||J(F_{:2})||_{\max}^2 & ||A||_F \leq n||A||_{\max} & \quad (104) \\
&\leq 4||J(F^R)||_{\max}^2 & \text{because the principal submatrix just removes entries} & \quad (105) \\
& & \text{and } J(F_{:2}) = J^{(2)}(F^R). & \\
&\leq 4||J(F^R)||_2^2 & ||A||_{\max} \leq ||A||_2 & \quad (106) \\
&\leq 4||J(F)||_2^2 & \text{Lemma 3} & \quad (107)
\end{aligned}$$

$\square$

**Lemma 5.** $||J^2(F')||_{2,1} \leq 4||J^2(F)||_2 = 4\gamma$.

*Proof.*

$$\begin{align}
||J^2(F')||_{2,1} &= ||J^2(F_{:2})||_{2,1} & \text{because } J^2(F') \text{ is } J^2(F_{:2}) \text{ augmented with zeros} \quad (108)\\
&= \sum_j ||J^2(F_{:2})_{\cdot j}||_2 & \text{by definition} \quad (109)\\
&\leq \sum_j ||J^2(F_{:2})_{\cdot j}||_1 & ||v||_2 \leq ||v||_1 \quad (110)\\
&= \sum_{ij} |J^2(F_{:2})_{ij}| & (111)\\
&\leq 4 \max_{ij} |J^2(F_{:2})_{ij}| & \text{because } J^2(F_{:2}) \text{ is a } 2 \times 2 \text{ matrix} \quad (112)\\
&= 4||J^2(F_{:2})_{ij}||_{\max} & \text{by definition} \quad (113)\\
&\leq 4||J^2(F^R)||_{\max}^2 & \text{because the principal submatrix just removes entries} \quad (114)\\
&\leq 4||J^2(F)||_2^2 & \text{Lemma 3} \quad (115)
\end{align}$$

□

**Theorem 6.** $||\nabla \times F'(x)||_2 \leq \sqrt{8(\beta^2 + L\gamma)}$

*Proof.*

$$\begin{align}
||\nabla \times F'(x)||_2^2 &\leq ||\nabla||_2^2 ||F'(x)||_2^2 & ||a \times b|| = ||a||||b|| \sin\theta \quad (116)\\
&= \Delta ||F'(x)||_2^2 & ||\nabla||_2^2 = \nabla \cdot \nabla = \Delta \quad (117)\\
&= \Delta\Big( \sum_{i=1}^{3} F_i'(x)^2 \Big) & ||F'(x)||_2^2 = F'(x) \cdot F'(x) \quad (118)\\
&= \sum_{i=1}^{3} \Delta\Big( F_i'(x)^2 \Big) & \text{linearity of } \nabla^2 \quad (119)\\
&= 2 \sum_{i=1}^{3} \sum_{j=1}^{3} \Big[ \frac{\partial F_i'(x)}{\partial x_j}^2 + F_i'(x) \frac{\partial^2 F_i'(x)}{\partial x_j^2} \Big] & \text{evaluating } \Delta \quad (120)\\
&= 2 \Big[ ||J(F'(x))||_F^2 + \sum_{j=1}^{3}\sum_{i=1}^{3} F_i'(x) \frac{\partial^2 F_i'(x)}{\partial x_j^2} \Big] & \text{by definition} \quad (121)\\
&= 2 \Big[ ||J(F'(x))||_F^2 + \sum_{j=1}^{3} \langle F'(x), \frac{\partial^2 F'(x)}{\partial x_j^2} \rangle \Big] & \text{write as inner product} \quad (122)\\
&\leq 2 \Big[ ||J(F'(x))||_F^2 + \sum_{j=1}^{3} ||F'(x)||_2 ||\frac{\partial^2 F'(x)}{\partial x_j^2}||_2 \Big] & a \cdot b = ||a||||b|| \cos\theta \quad (123)\\
&= 2 \Big[ ||J(F'(x))||_F^2 + ||F'(x)||_2 \sum_{j=1}^{3} ||\frac{\partial^2 F'(x)}{\partial x_j^2}||_2 \Big] & \text{pull in sum} \quad (124)\\
&= 2 \Big[ ||J(F'(x))||_F^2 + ||F'(x)||_2 ||J^2(F'(x))||_{2,1} \Big] & \text{by definition} \quad (125)\\
&\leq 2(4\beta^2 + 4L\gamma) & \text{by Lemmas 3, 4, and 5.} \quad (126)
\end{align}$$

□

## A.9 Algorithmic Game Theory: A Venn Diagram

Here, we consider cost-minimization games where $C_i(\mathbf{s})$ is player $i$'s cost function and $C(\mathbf{s}) = \sum_{i=1}^{K} C_i(\mathbf{s})$. Player $i$'s strategy set is $\mathbf{s}_i$ and $\mathbf{s}_{-i}$ represents the strategy sets of all players except player $i$.

**Definition 3** (Smooth Game). *A cost-minimization game is ($\lambda,\mu$)-smooth if for every two outcomes $\mathbf{s}$ and $\mathbf{s}^*$,*

$$\sum_{i=1}^{K} C_i(\mathbf{s}_i^*, \mathbf{s}_{-i}) \leq \lambda \cdot C(\mathbf{s}^*) + \mu \cdot C(\mathbf{s}). \tag{127}$$

**Definition 4** (Convex Game). *A cost-minimization game is convex if $C_i(\mathbf{s}_i, \mathbf{s}_{-i})$ is convex in $\mathbf{s}_i$ $\forall i$.*

**Definition 5** (Monotone Game). *A cost-minimization game is monotone if the game dynamics are monotone. Here, we assume all players are running OMoMD.*

$$F = \begin{pmatrix} \nabla_{\mathbf{s}_0} C_0 \\ ... \\ \nabla_{\mathbf{s}_K} C_K \end{pmatrix} \tag{128}$$

*Monotonicity requires that the symmetrized Jacobian of $F$ be positive semidefinite: $J + J^T \succeq 0$.*

**Definition 6** (Socially-Convex Game). *A cost-minimization game is socially-convex if*

1. *There exists $\lambda_i > 0$ such that $\sum_{i=1}^{K} \lambda_i = 1$, $g(\mathbf{s}) = \sum_{i=1}^{K} \lambda_i C_i(\mathbf{s})$ is convex in $\mathbf{s}$, and*
2. *$C_i(\mathbf{s}_i, \mathbf{s}_{-i})$ is concave in $\mathbf{s}_{-i}$ $\forall i$.*

*The definition was originally written for concave.*

**Theorem 7** (Monotone $\Longrightarrow$ Convex). *If a game is monotone, it is also convex.*

*Proof.* For each player $i$, we show that $C_i(\mathbf{s}_i, \mathbf{s}_{-i})$ is convex in $\mathbf{s}_i$ for any fixed $\mathbf{s}_{-i}$ (i.e., $\mathbf{s}_{-i} = \mathbf{s}'_{-i}$). The associated map is given by (128). Let $K_i := \{\mathbf{s}, \mathbf{s}' \in K \; s.t. \; \mathbf{s}_{-i} = \mathbf{s}'_{-i}\}$. Starting with the definition of monotonicity, we have

$$\langle F(\mathbf{s}) - F(\mathbf{s}'), \mathbf{s} - \mathbf{s}'\rangle \geq 0 \; \forall \; \mathbf{s}, \mathbf{s}' \in K \tag{129}$$

$$\Longrightarrow \langle F(\mathbf{s}) - F(\mathbf{s}'), \mathbf{s} - \mathbf{s}'\rangle \geq 0 \forall \; \mathbf{s}, \mathbf{s}' \in K_i \tag{130}$$

$$= \sum_j \langle F_j(\mathbf{s}) - F_j(\mathbf{s}'), \mathbf{s}_j - \mathbf{s}'_j\rangle \geq 0 \; \forall \; \mathbf{s}, \mathbf{s}' \in K_i \tag{131}$$

$$= \sum_{j \neq i} \langle F_j(\mathbf{s}) - F_j(\mathbf{s}'), \underbrace{\mathbf{s}_j - \mathbf{s}'_j}_{0}\rangle \geq 0 \; \forall \; \mathbf{s}, \mathbf{s}' \in K_i \tag{132}$$

$$+ \langle F_i(\mathbf{s}) - F_i(\mathbf{s}'), \mathbf{s}_i - \mathbf{s}'_i\rangle \geq 0 \; \forall \; \mathbf{s}, \mathbf{s}' \in K_i \tag{133}$$

$$= \langle \nabla C_i(\mathbf{s}) - \nabla C_i(\mathbf{s}'), \mathbf{s}_i - \mathbf{s}'_i\rangle \geq 0 \; \forall \; \mathbf{s}, \mathbf{s}' \in K_i \tag{134}$$

which is the definition of convexity, so $C_i$ is convex in $\mathbf{s}_i$. $\square$

**Theorem 8** (Socially-Convex $\Longrightarrow$ Convex). *Lemma 2.2 in (Even-Dar, Mansour, and Nadav 2009) with convex swapped for concave.*

**Theorem 9** (Socially-Convex $\Longrightarrow$ $\lambda-$Monotone). *A game that is socially-convex with parameters $\lambda$ implies a scaling of the game with the same parameters that is monotone (credit to Peng Shi).*

*Proof.* Let $C'_i = \lambda_i C_i$ and let $J'$ be the Jacobian of the map, $F'$, corresponding to $C'_i$ (128). In addition, define the following matrices

1. $D$ such that $D_{ii} = \lambda_i \frac{\partial^2 C_i}{\partial \mathbf{s}_i^2}$ and $D_{ij} = 0$ $\forall i \neq j$.

2. $G^k$ is such that $\forall i \; G^k_{ik} = G^k_{ki} = 0$ and $\forall i \neq k, j \neq k$, $G^k_{ij} = \lambda_k \frac{\partial^2 C_k}{\partial \mathbf{s}_i \mathbf{s}_j}$.

3. $H$ is the Hessian of $g(\mathbf{s}) = \sum_k \lambda_k C_k(\mathbf{s})$ (i.e., $H_{ij} = \sum_k \lambda_k \frac{\partial^2 C_k}{\partial \mathbf{s}_i \mathbf{s}_j}$).

Note that $D$, $-G^k$, and $H$ are all positive semidefinite matrices. This follows from the fact that player costs are convex in their own strategies, concave in other players' strategies, and the socially-convex condition respectively.

Continuing the proof, for every $i \neq j$,

$$(D - \sum_k G^k + H)_{ij} = 0 - \sum_{k \neq i,j} \lambda_k \frac{\partial^2 C_k}{\partial \mathbf{s}_i \mathbf{s}_j} + \sum_k \lambda_k \frac{\partial^2 C_k}{\partial \mathbf{s}_i \mathbf{s}_j} \tag{135}$$

$$= \lambda_i \frac{\partial^2 C_i}{\partial \mathbf{s}_i \mathbf{s}_j} + \lambda_j \frac{\partial^2 C_j}{\partial \mathbf{s}_i \mathbf{s}_j} \tag{136}$$

$$= (J' + J'^T)_{ij}. \tag{137}$$

Moreover, for every $i = j$,

$$(D - \sum_k G^k + H)_{ii} = \frac{\partial^2 C_i}{\partial \mathbf{s}_i^2} - \sum_{k \neq i} \lambda_k \frac{\partial^2 C_k}{\partial \mathbf{s}_i^2} + \sum_k \lambda_k \frac{\partial^2 C_k}{\partial \mathbf{s}_i^2} \tag{138}$$

$$= 2\lambda_i \frac{\partial^2 C_i}{\partial \mathbf{s}_i^2} \tag{139}$$

$$= (J' + J'^T)_{ii}. \tag{140}$$

Therefore, $(J + J'^T) = D - \sum_k G^k + H$. Each of the matrices $(D, G^k, H)$ is positive semidefinite, therefore $(J + J'^T) \succeq 0$, hence $F'$ is monotone. $\square$

**a. Smooth** The following cost-minimization game is $(\frac{1}{2}, \frac{1}{2})$-smooth.

$$C_1 = C_2 = -\cos(r) - \cos(c) \tag{141}$$

*Proof.*

$$\sum_{i=1}^{K} C_i(s_i^*, \mathbf{s}_{-i}) = -\cos(r^*) - \cos(c) - \cos(r) - \cos(c^*) \tag{142}$$

$$\leq \lambda \cdot C(\mathbf{s}^*) = -\cos(r^*) - \cos(c^*) \tag{143}$$
$$+ \mu \cdot C(\mathbf{s}) = -\cos(r) - \cos(c) \tag{144}$$
$$= -\cos(r^*) - \cos(c) - \cos(r) - \cos(c^*) \tag{145}$$

$\square$

This game is not convex, therefore, it is neither monotone nor socially-convex.

**b. Smooth, Convex** Consider the following cost-minimization game.

$$C_1 = r^2(\sin(c) + 1.25) \tag{146}$$
$$C_2 = c^2(\sin(r) + 1.25) \tag{147}$$

This game is $(10, 0)$-smooth.

*Proof.*

$$\sum_{i=1}^{K} C_i(s_i^*, \mathbf{s}_{-i}) = r^{*2}(\sin(c) + 1.25) + c^{*2}(\sin(r) + 1.25) \tag{148}$$

$$\leq 2.25(r^{*2} + c^{*2}) \tag{149}$$
$$\lambda \cdot C(\mathbf{s}^*) = 10[r^{*2}(\sin(c^*) + 1.25) + c^{*2}(\sin(r^*) + 1.25)] \tag{150}$$
$$\geq 2.5(r^{*2} + c^{*2}) \tag{151}$$
$$2.25(r^{*2} + c^{*2}) \leq 2.5(r^{*2} + c^{*2}) \tag{152}$$

$\square$

Clearly, $C_1$ is convex in $r$ and $C_2$ is convex in $c$, therefore the game is convex.
The corresponding map is not monotone.

*Proof.*

$$F = \begin{pmatrix} 2r(\sin(c)+1.25) \\ 2c(\sin(r)+1.25) \end{pmatrix} \tag{153}$$

$$J = 2\begin{pmatrix} \sin(c)+1.25 & r\cos(c) \\ c\cos(r) & \sin(r)+1.25 \end{pmatrix} \tag{154}$$

$$J_s = 2\begin{pmatrix} \sin(c)+1.25 & \frac{r}{2}\cos(c)+\frac{c}{2}\cos(r) \\ \frac{r}{2}\cos(c)+\frac{c}{2}\cos(r) & \sin(r)+1.25 \end{pmatrix} \tag{155}$$

$$J_s\Big|_{r=c=-\frac{\pi}{4}} = 2\begin{pmatrix} -\frac{\sqrt{2}}{2}+1.25 & -\frac{\pi}{4}\cos(-\frac{\pi}{4}) \\ -\frac{\pi}{4}\cos(-\frac{\pi}{4}) & -\frac{\sqrt{2}}{2}+1.25 \end{pmatrix} \not\succeq 0 \tag{156}$$

$\square$

$C_1$ is not concave with respect to $c$. Likewise, $C_2$ is not concave with respect to $r$. Therefore, this game is not socially-convex.

**c. Smooth, Convex, Monotone**  Consider the following cost-minimization game.
$$C_1 = r^2 + c^2 \tag{157}$$
$$C_2 = r^2 + c^2 \tag{158}$$

This game is $(\frac{1}{2}, \frac{1}{2})$-smooth.

*Proof.*
$$\sum_{i=1}^{K} C_i(s_i^*, \mathbf{s}_{-i}) = r^{*2} + c^2 + r^2 + c^{*2} \tag{159}$$
$$\lambda \cdot C(\mathbf{s}^*) = r^{*2} + c^{*2} \tag{160}$$
$$\mu \cdot C(\mathbf{s}) = r^2 + c^2 \tag{161}$$
$$r^{*2} + c^2 + r^2 + c^{*2} \leq r^{*2} + c^2 + r^2 + c^{*2} \tag{162}$$
□

Clearly, $C_1$ is convex in $r$ and $C_2$ is convex in $c$, therefore the game is convex.

The corresponding map is monotone.

*Proof.*
$$F = \begin{pmatrix} 2r \\ 2c \end{pmatrix} \tag{163}$$
$$J = J_s = \begin{pmatrix} 2 & 0 \\ 0 & 2 \end{pmatrix} \succeq 0 \tag{164}$$
□

$C_1$ is not concave with respect to $c$. Likewise, $C_2$ is not concave with respect to $r$. Therefore, this game is not socially-convex.

**d. Smooth, Convex, Socially-Convex**  Consider the following cost-minimization game (inspired by modified Tail Drop policy in routing networks) over $(r, c) \in (0, 1]^2 = K$.
$$C_1 = -\frac{1}{2}(\frac{r}{r+c}) \tag{165}$$
$$C_2 = -\frac{c}{r+c} \tag{166}$$

This game is $(\frac{1}{2}, -1)$-smooth.

*Proof.*
$$\sum_{i=1}^{K} C_i(s_i^*, \mathbf{s}_{-i}) = -\frac{1}{2}(\frac{r^*}{r^* + c}) - \frac{c^*}{r + c^*} \tag{167}$$
$$\leq 0 \text{ over } K \tag{168}$$
$$\mu \cdot C(\mathbf{s}^*) = 1 - \frac{1}{2}(\frac{r}{r+c}) \geq \frac{1}{2} \tag{169}$$
$$\lambda \cdot C(\mathbf{s}^*) = -\frac{1}{2}(1 - \frac{1}{2}(\frac{r}{r+c})) \geq -\frac{1}{2} \tag{170}$$
$$\sum_{i=1}^{K} C_i(s_i^*, \mathbf{s}_{-i}) \leq 0 \leq \mu \cdot C(\mathbf{s}^*) + \lambda \cdot C(\mathbf{s}^*) \tag{171}$$
□

$C_1$ is convex in $r$ over $K$ and $C_2$ is convex in $c$ over $K$, therefore the game is convex.

*Proof.*
$$\frac{\partial^2 C_1}{\partial r^2} = \frac{c}{(r+c)^3} \geq 0 \text{ over } K \tag{172}$$
$$\frac{\partial^2 C_2}{\partial c^2} = \frac{2r}{(r+c)^3} \geq 0 \text{ over } K \tag{173}$$
□

The corresponding map is not monotone.

*Proof.*

$$F = -\frac{1}{(r+c)^2}\begin{pmatrix}\frac{c}{2}\\\frac{c}{r}\end{pmatrix} \tag{174}$$

$$J = \frac{1}{(r+c)^3}\begin{pmatrix}c & \frac{c-r}{2}\\r-c & 2r\end{pmatrix} \tag{175}$$

$$J_s = \frac{1}{(r+c)^3}\begin{pmatrix}c & \frac{r-c}{4}\\\frac{r-c}{4} & 2r\end{pmatrix} \tag{176}$$

$$Det(J_s) = 2rc - \frac{1}{16}(r-c)^2 \tag{177}$$

$$Det(J_s)|_{r=0.01, c=1} = -0.041 \le 0 \tag{178}$$

The determinant of $J_s$ is negative over a subset of the domain (e.g., $r \le 0.01, c = 1$), therefore, $J_s$ is not positive semidefinite. Hence, $F$ is not monotone. □

Let $\lambda_1 = \frac{2}{3}$ and $\lambda_2 = \frac{1}{3}$. Then $\lambda_1 C_1 + \lambda_2 C_2 = -\frac{1}{3}$, which is convex in $(r, c)$. Also, $C_1$ is concave with respect to $c$ and $C_2$ is concave with respect to $r$, therefore, this game is socially-convex.

*Proof.*

$$\frac{\partial^2 C_1}{\partial c^2} = -\frac{r}{(r+c)^3} \le 0 \text{ over } K \tag{179}$$

$$\frac{\partial^2 C_2}{\partial r^2} = -\frac{2c}{(r+c)^3} \le 0 \text{ over } K \tag{180}$$

□

**e. Smooth, Convex, Monotone, Socially-Convex** Consider the following cost-minimization game.

$$C_1 = r \tag{181}$$
$$C_2 = c \tag{182}$$

This game is $(1, 0)$-smooth.

*Proof.*

$$\sum_{i=1}^{K} C_i(s_i^*, \mathbf{s}_{-i}) = r^* + c^* \tag{183}$$

$$\lambda \cdot C(\mathbf{s}^*) = r^* + c^* \tag{184}$$

$$r^* + c^* \le r^* + c^* \tag{185}$$

□

Clearly, $C_1$ is convex in $r$ and $C_2$ is convex in $c$, therefore the game is convex.

The corresponding map is monotone.

*Proof.*

$$F = \begin{pmatrix}1\\1\end{pmatrix} \tag{186}$$

$$J = J_s = \begin{pmatrix}0 & 0\\0 & 0\end{pmatrix} \succeq 0 \tag{187}$$

□

Let $\lambda_1 = \lambda_2 = \frac{1}{2}$. Then $\lambda_1 C_1 + \lambda_2 C_2 = \frac{1}{2}(r+c)$, which is convex in $(r, c)$. Also, $C_1$ is concave with respect to $c$ and $C_2$ is concave with respect to $r$, therefore, this game is socially-convex.

**f. Convex** Consider the following cost-minimization game.

$$C_1 = r^2 + \frac{r}{c^2 + \frac{1}{4}} - \frac{9}{5}c \tag{188}$$

$$C_2 = c^2 + \frac{c}{r^2 + \frac{1}{4}} - \frac{9}{5}r \tag{189}$$

This game is not smooth.

*Proof.*

$$\sum_{i=1}^{K} C_i(s_i^*, \mathbf{s}_{-i}) = r^* + c^* \tag{190}$$

$$\lambda \cdot C(\mathbf{s}^*) = r^* + c^* \tag{191}$$

$$r^* + c^* \leq r^* + c^* \tag{192}$$

$\square$

$C_1$ is convex in $r$ and $C_2$ is convex in $c$, therefore the game is convex.
The corresponding map is not monotone.

*Proof.*

$$F = \begin{pmatrix} 2r + \frac{1}{c^2 + \frac{1}{4}} \\ 2c + \frac{1}{r^2 + \frac{1}{4}} \end{pmatrix} \tag{193}$$

$$J = J_s = \begin{pmatrix} 2 & -\frac{2c}{(c^2 + \frac{1}{4})^2} \\ -\frac{2r}{(r^2 + \frac{1}{4})^2} & 2 \end{pmatrix} \tag{194}$$

$$J_s \Big|_{r=c=\frac{1}{4}} = \begin{pmatrix} 2 & -5.12 \\ -5.12 & 2 \end{pmatrix} \not\succeq 0 \tag{195}$$

$\square$

This game is not socially-convex because $C_1$ is not concave with respect to $c$ and likewise for $C_2$ and $r$. For example, $C_1(r = 1, c) = 1 + \frac{1}{c^2 + \frac{1}{4}} - \frac{9}{5}c$ is not concave with respect to $c$.

**g. Convex, Monotone** Consider the following cost-minimization game.

$$C_1 = r^2 + c^2 - 2 \tag{196}$$

$$C_2 = r^2 + c^2 + r + c - 2 \tag{197}$$

This game is not smooth.

*Proof.* Consider $(r, c) = (1, -1)$ and $(r^*, c^*) = (-1, 1)$.

$$\sum_{i=1}^{K} C_i(s_i^*, \mathbf{s}_{-i}) = 2 \tag{198}$$

$$\mu \cdot C(\mathbf{s}) = 0 \tag{199}$$

$$\lambda \cdot C(\mathbf{s}^*) = 0 \tag{200}$$

$$2 \not\leq 0 \tag{201}$$

$\square$

$C_1$ is convex in $r$ and $C_2$ is convex in $c$, therefore the game is convex.
The corresponding map is monotone.

*Proof.*

$$F = \begin{pmatrix} 2r \\ 2c + 1 \end{pmatrix} \tag{202}$$

$$J = J_s = \begin{pmatrix} 2 & 0 \\ 0 & 2 \end{pmatrix} \succeq 0 \tag{203}$$

$\square$

This game is not socially-convex because $C_1$ is not concave with respect to $c$ and likewise for $C_2$ and $r$.

**h. Convex, Socially-Convex** Consider the following cost-minimization game (inspired by modified Tail Drop policy in routing networks) over $(r, c) \in (0, 1]^2 = K$.

$$C_1 = -\frac{1}{2}\left(\frac{r}{r+c}\right) + \frac{3}{4} \tag{204}$$

$$C_2 = -\frac{c}{r+c} \tag{205}$$

This game is not smooth.

*Proof.* Consider $(r, c) = (1, 1)$ and $(r^*, c^*) = (\frac{1}{2}, \frac{1}{2})$.

$$\sum_{i=1}^{K} C_i(s_i^*, \mathbf{s}_{-i}) = \frac{1}{4} \tag{206}$$

$$\mu \cdot C(\mathbf{s}^*) = 0 \tag{207}$$

$$\lambda \cdot C(\mathbf{s}^*) = 0 \tag{208}$$

$$\frac{1}{4} \not\leq 0 \tag{209}$$

$\square$

$C_1$ is convex in $r$ over $K$ and $C_2$ is convex in $c$ over $K$, therefore the game is convex.

*Proof.*

$$\frac{\partial^2 C_1}{\partial r^2} = \frac{c}{(r+c)^3} \geq 0 \text{ over } K \tag{210}$$

$$\frac{\partial^2 C_2}{\partial c^2} = \frac{2r}{(r+c)^3} \geq 0 \text{ over } K \tag{211}$$

$\square$

The corresponding map is not monotone.

*Proof.*

$$F = -\frac{1}{(r+c)^2} \begin{pmatrix} \frac{c}{2} \\ r \end{pmatrix} \tag{212}$$

$$J = \frac{1}{(r+c)^3} \begin{pmatrix} c & \frac{c-r}{2} \\ r-c & 2r \end{pmatrix} \tag{213}$$

$$J_s = \frac{1}{(r+c)^3} \begin{pmatrix} c & \frac{r-c}{4} \\ \frac{r-c}{4} & 2r \end{pmatrix} \tag{214}$$

$$Det(J_s) = 2rc - \frac{1}{16}(r-c)^2 \tag{215}$$

$$Det(J_s)|_{r=0.01, c=1} = -0.041 \leq 0 \tag{216}$$

The determinant of $J_s$ is negative over a subset of the domain (e.g., $r \leq 0.01, c = 1$), therefore, $J_s$ is not positive semidefinite. Hence, $F$ is not monotone. $\square$

Let $\lambda_1 = \frac{2}{3}$ and $\lambda_2 = \frac{1}{3}$. Then $\lambda_1 C_1 + \lambda_2 C_2 = \frac{1}{6}$, which is convex in $(r, c)$. Also, $C_1$ is concave with respect to $c$ and $C_2$ is concave with respect to $r$, therefore, this game is socially-convex.

*Proof.*

$$\frac{\partial^2 C_1}{\partial c^2} = -\frac{r}{(r+c)^3} \leq 0 \text{ over } K \tag{217}$$

$$\frac{\partial^2 C_2}{\partial r^2} = -\frac{2c}{(r+c)^3} \leq 0 \text{ over } K \tag{218}$$

$\square$

**i. Convex, Monotone, Socially-Convex**  Consider the following cost-minimization game.

$$C_1 = r^2 - 1 \tag{219}$$
$$C_2 = c^2 + r + c - 1 \tag{220}$$

This game is not smooth.

*Proof.* Consider $(r, c) = (1, -1)$ and $(r^*, c^*) = (-1, 1)$.

$$\sum_{i=1}^{K} C_i(s_i^*, \mathbf{s}_{-i}) = 2 \tag{221}$$
$$\mu \cdot C(\mathbf{s}) = 0 \tag{222}$$
$$\lambda \cdot C(\mathbf{s}^*) = 0 \tag{223}$$
$$2 \not\leq 0 \tag{224}$$

□

$C_1$ is convex in $r$ and $C_2$ is convex in $c$, therefore the game is convex.
The corresponding map is monotone.

*Proof.*

$$F = \begin{pmatrix} 2r \\ 2c+1 \end{pmatrix} \tag{225}$$
$$J = J_s = \begin{pmatrix} 2 & 0 \\ 0 & 2 \end{pmatrix} \succeq 0 \tag{226}$$

□

Let $\lambda_1 = \lambda_2 = \frac{1}{2}$. Then $\lambda_1 C_1 + \lambda_2 C_2 = \frac{1}{2}(r^2 + c^2 + r + c - 2)$, which is convex in $(r, c)$. Also, $C_1$ is concave with respect to $c$ and $C_2$ is concave with respect to $r$, therefore, this game is socially-convex.

## A.10  Concave Games

Several well known concave games are monotone.

**Linear Cournot Competition**  In linear *Cournot* competition, $N$ firms compete for customers by adjusting the quantity of goods they produce, $x_i$. Firms pay a cost for producing those goods, $c_i(x_i)$, which is assumed to be a convex function in $x_i$. The prices for goods are set by the consumer demand markets according to a price function, $p(x) = a - b \sum_k x_k$, with $a, b > 0$. The firms attempt to maximize their utility or profit functions, $u_i(x) = x_i p(x) - c_i(x_i)$. Here, we show that the map associated with the game, $F(x) = \{-\frac{\partial u_0}{\partial x_0}, \ldots, -\frac{\partial u_N}{\partial x_N}\}$, is monotone.

First we derive the first and second partial derivatives.

$$\frac{\partial u_i}{\partial x_i} = p(x) - bx_i - \frac{\partial c_i}{\partial x_i} \tag{227}$$
$$\frac{\partial^2 u_i}{\partial x_i^2} = -2b - \frac{\partial^2 c_i}{\partial x_i^2} \tag{228}$$
$$\frac{\partial^2 u_i}{\partial x_i x_j} = -b \tag{229}$$

These derivatives, in turn, define the Jacobian, $Jac(F)$, which can be decomposed into a constant matrix with all entires equal to $b$ and a diagonal matrix consisting of $b + \frac{\partial^2 c_i}{\partial x_i^2}$. A constant matrix with positive entries $b$ is rank-1 with eigenvalues $\{Nb\} + \{0\}^{N-1}$. The cost functions, $c_i$, are assumed to be convex, therefore, the diagonal matrix is positive-definite. This implies that the sum of the two *symmetric* matrices is positive definite. It follows that $F$ is monotone.

Let $v(t) = o + (x - o)t$ and $dv = (x - o)dt$.

$$\int_{v:o \to x} \langle -F(v), dv \rangle = \int_0^1 \sum_i (a - bv_i(t) - b\sum_k v_k(t))(x_i - o_i) dt - \sum_i (c_i(x) - c_i(o)) \tag{230}$$

$$= \sum_i (x_i - o_i) \int_0^1 (a - bv_i(t) - b\sum_k v_k(t)) dt - \sum_i c_i(x) \tag{231}$$

$$= \sum_i (x_i - o_i) \int_0^1 (a - bo_i - b(x_i - o_i)t - b\sum_k o_k + (x_k - o_k)t) dt - \sum_i (c_i(x) - c_i(o)) \tag{232}$$

$$= \sum_i (x_i - o_i)(at - bo_i t - b(x_i - o_i)\frac{t^2}{2} - b\sum_k o_k t + (x_k - o_k)\frac{t^2}{2})\Big|_0^1 - \sum_i (c_i(x) - c_i(o)) \tag{233}$$

$$= \sum_i (x_i - o_i)(a - bo_i - \frac{b}{2}(x_i - o_i) - b\sum_k o_k + \frac{1}{2}(x_k - o_k)) - \sum_i (c_i(x) - c_i(o)) \tag{234}$$

$$= \sum_i (x_i - o_i)(a - b\frac{o_i + x_i}{2} - b\sum_k \frac{o_k + x_k}{2}) - \sum_i (c_i(x) - c_i(o)) \tag{235}$$

$$= \sum_i x_i p(z_i) - c_i(x) - \sum_i o_i p(z_i) - c_i(o) \quad \text{where } z_i = \frac{1}{2}(o_i + x_i + \sum_k o_k + x_k) \tag{236}$$

So *auto*-welfare is calculating profits with player specific prices. Specifically, each player's price is set as a deviation from the average supply of $o$ and $x$. If $o$ is set to the origin, $z_i$ is half the total market supply except with player $i$'s supply at full. More generally, if player $i$ chooses to flood the market with good, $x_i$, *auto*-welfare computes its contribution to the sum with a lower price point.

**Linear Resource Allocation** In a resource allocation game, $N$ users share a communication channel with finite capacity (e.g., $C = 1$). Each user $i$ submits a bid, $x_i \in [\epsilon > 0, 1]$, to the communication network which then allocates a fraction of the communication channel to each user according an allocation function, $M_i(x) = x_i / \sum_k x_k$. Each user plays to maximize its utility, $u_i(x) = \psi_i(M_i(x)) - \alpha_i x_i$, with $\alpha_i > 0$. Here, we consider a simplified value function, $\psi_i(z) = \beta z$, with $\beta > 0$ and show that the map associated with the game, $F(x) = \{-\frac{\partial u_0}{\partial x_0}, \ldots, -\frac{\partial u_N}{\partial x_N}\}$, is monotone.

First we derive the first and second partial derivatives.

$$\frac{\partial M_i(x)}{\partial x_i} = \frac{1}{\sum_k x_k}\left[1 - \frac{x_i}{\sum_k x_k}\right] \tag{237}$$

$$\frac{\partial M_i(x)}{\partial x_j} = \frac{1}{\sum_k x_k}\left[0 - \frac{x_i}{\sum_k x_k}\right] \tag{238}$$

$$\frac{\partial^2 M_i(x)}{\partial x_i^2} = -\frac{1}{(\sum_k x_k)^2}\left[2 - \frac{2x_i}{\sum_k x_k}\right] \tag{239}$$

$$\frac{\partial^2 M_i(x)}{\partial x_i x_j} = -\frac{1}{(\sum_k x_k)^2}\left[1 - \frac{2x_i}{\sum_k x_k}\right] \tag{240}$$

$$\frac{\partial^2 u_i}{\partial x_i^2} = \beta \frac{\partial^2 M_i(x)}{\partial x_i^2} \tag{241}$$

$$\frac{\partial^2 u_i}{\partial x_i x_j} = \beta \frac{\partial^2 M_i(x)}{\partial x_i x_j} \tag{242}$$

These derivatives, in turn, define the Jacobian, $Jac(F)$, which can be decomposed into a rank-1 matrix, $M$, with constant rows and an identity matrix, $\mathbb{I}_N$. Let $z_i = x_i / \sum_k x_k \in (0, 1]$.

$$Jac(F)_{ij} = \underbrace{\frac{\beta}{(\sum_k x_k)^2}}_{\geq 0}\Big[\underbrace{1 - 2z_i}_{M_{ij}} + \underbrace{\mathbb{I}(i = j)}_{\mathbb{I}_N}\Big] \tag{243}$$

We can prove $Jac(F)$ is monotone by showing $\frac{1}{2}(Jac(F) + Jac(F)^T) \succeq 0$. As a first step, we'll lower bound the eigenvalues of a symmetrized $M$.

$$M^{(s)}_{ij} = \frac{1}{2}(M + M^T)_{ij} = 1 - (z_i + z_j) \text{ is at most a rank-2 matrix} \tag{244}$$

$$\implies \lambda(M^{(s)}) = \{\lambda_{lo}, \lambda_{hi}\} + \{0\}^{N-2} \tag{245}$$

$$\text{Tr}(M^{(s)}) = \lambda_{lo} + \lambda_{hi} = N - 2 \tag{246}$$

$$||M^{(s)}||_1 = ||M^{(s)}||_\infty = |1 - 2z_i| + \sum_{j \neq i} |1 - (z_i + z_j)| \tag{247}$$

$$= |1 - 2z_i| + \sum_{j \neq i} 1 - (z_i + z_j) \qquad z_i, z_j > \epsilon, \sum_k z_k = 1 \implies z_i + z_j < 1 \tag{248}$$

$$= |1 - 2z_i| + (N - 2)(1 - z_i) \leq N - 1 \tag{249}$$

$$\rho(M^{(s)}) = ||M^{(s)}||_2 \leq \sqrt{||M^{(s)}||_1 ||M^{(s)}||_\infty} = N - 1 \qquad \textit{Holder's inequality} \tag{250}$$

Assume $\min \lambda(M^{(s)}) = \lambda_{lo} < -1 \implies \max \lambda(M^{(s)}) = \lambda_{hi} > N - 1$ contradicts $\rho(M^{(s)})$ \qquad (251)

$$\implies \min \lambda(M^{(s)}) \geq -1 \tag{252}$$

The eigenvalues of $M^{(s)}$ are lower bounded by $-1$, therefore, the eigenvalues of the sum of $M^{(s)}$ and an identity matrix are lower bounded by 0.

$$J^{(s)} = \frac{1}{2}(Jac(F) + Jac(F)^2) = \underbrace{\frac{\beta}{(\sum_k x_k)^2}}_{\geq 0} \left[ M^{(s)} + \mathbb{1}_N \right] \tag{253}$$

$$\implies \min \lambda(J^{(s)}) \geq 0 \tag{254}$$

$$\implies F \text{ is monotone } \checkmark \tag{255}$$

Next, we'll compare welfare and *auto*-welfare. For the moment, consider welfare with $\alpha_i = 0$:

$$W = \sum_i u_i = \beta \sum_i \frac{x_i}{\sum_k x_k} = \beta. \tag{256}$$

Notice that without $\alpha_i$, welfare regret, $\sum_t W(x_t) - W(u_T)$, is identically zero and is a pointless quantity to maximize. If we include $\alpha_i$,

$$W = \sum_i u_i = \beta - \sum_i \alpha_i x_i, \tag{257}$$

which has a simple closed-form solution for any $\alpha_i > 0$: $x_i = \epsilon$. This result is independent of the parameters of the utility functions, $\beta$ and $\alpha_i$. Next, we'll compute *auto*-welfare to contrast. Let $v(t) = o + (x - o)t$ and $dv = (x - o)dt$. The critical component of *auto*-welfare is the path integral.

$$\int_{v:o \to x} \langle -F(v), dv \rangle = \int_0^1 \sum_i \left( \frac{\beta}{\sum_k v_k(t)} \left[ 1 - \frac{v_i(t)}{\sum_k v_k(t)} \right] - \alpha_i \right)(x_i - o_i) dt \tag{258}$$

$$= \sum_i \int_0^1 \frac{\beta}{\sum_k v_k(t)} \left[ 1 - \frac{v_i(t)}{\sum_k v_k(t)} \right](x_i - o_i) dt - \sum_i \int_0^1 \alpha_i (x_i - o_i) dt \tag{259}$$

$$= \sum_i \int_0^1 \frac{\beta(x_i - o_i)}{\sum_k o_k + (x_k - o_k)t} \left[ 1 - \frac{o_i + (x_i - o_i)t}{\sum_k o_k + (x_k - o_k)t} \right] dt - \sum_i \alpha_i (x_i - o_i) \tag{260}$$

$$= \sum_i \int_0^1 \frac{\beta(x_i - o_i)}{\sum_k o_k + t \sum_k (x_k - o_k)} \left[ 1 - \frac{o_i + t(x_i - o_i)}{\sum_k o_k + t \sum_k (x_k - o_k)} \right] dt - \sum_i \alpha_i (x_i - o_i) \tag{261}$$

$$= \sum_i \beta(x_i - o_i) \int_0^1 \frac{1}{s_o + t(s_x - s_o)} \left[ 1 - \frac{o_i + t(x_i - o_i)}{s_o + t(s_x - s_o)} \right] dt - \sum_i \alpha_i (x_i - o_i) \tag{262}$$

Breaking apart the left integrand, we first compute the following.

$$\int_0^1 \frac{1}{s_o + t(s_x - s_o)} dt = \frac{\ln(s_o + t(s_x - s_o))}{s_x - s_o} \bigg|_0^1 = \frac{\ln(s_x/s_o)}{s_x - s_o} \tag{263}$$

We'll use integration by parts on the other part. Let $u = o_i + t(x_i - o_i)$, $dv = (s_o + t(s_x - s_o))^{-2}dt$, $v = -\frac{1}{s_x - s_o}(s_o + t(s_x - s_o))^{-1}$, and $du = x_i - o_i dt$.

$$\int_0^1 \frac{o_i + t(x_i - o_i)}{(s_o + t(s_x - s_o))^2} dt = \int_0^1 u\, dv = uv\Big|_0^1 - \int_0^1 v\, du \tag{264}$$

$$uv\Big|_0^1 = -\frac{o_i + t(x_i - o_i)}{(s_x - s_o)(s_o + t(s_x - s_o))}\Big|_0^1 \tag{265}$$

$$= \frac{o_i}{s_o(s_x - s_o)} - \frac{x_i}{s_x(s_x - s_o))} \tag{266}$$

$$-\int_0^1 v\, du = \int_0^1 \frac{1}{s_x - s_o}(s_o + t(s_x - s_o))^{-1}(x_i - o_i)dt \tag{267}$$

$$= \frac{x_i - o_i}{s_x - s_o}\int_0^1 \frac{1}{s_o + t(s_x - s_o)} dt \tag{268}$$

$$= \frac{x_i - o_i}{s_x - s_o}\frac{\ln(s_x/s_o)}{s_x - s_o} \tag{269}$$

Combining the two results, we find

$$\int_{v:o \to x} \langle -F(v), dv \rangle = \sum_i \beta(x_i - o_i)\left[\frac{\ln(s_x/s_o)}{s_x - s_o}(1 - \frac{x_i - o_i}{s_x - s_o}) + \frac{x_i}{s_x(s_x - s_o)} - \frac{o_i}{s_o(s_x - s_o)}\right] - \sum_i \alpha_i(x_i - o_i) \tag{270}$$

$$= \sum_i \beta \frac{x_i - o_i}{s_x - s_o}\left[\ln(s_x/s_o)(1 - \frac{x_i - o_i}{s_x - s_o}) + \frac{x_i}{s_x} - \frac{o_i}{s_o}\right] - \sum_i \alpha_i(x_i - o_i) \tag{271}$$

$$= \beta \ln(s_x/s_o)\sum_i \frac{x_i - o_i}{s_x - s_o} - \beta \ln(s_x/s_o)\sum_i (\frac{x_i - o_i}{s_x - s_o})^2 + \beta \sum_i \frac{x_i - o_i}{s_x - s_o}\left[\frac{x_i}{s_x} - \frac{o_i}{s_o}\right] - \sum_i \alpha_i(x_i - o_i) \tag{272}$$

$$= \beta \ln(s_x/s_o)\left(1 - \frac{||x - o||_2^2}{(s_x - s_o)^2}\right) + \beta \sum_i \frac{x_i - o_i}{s_x - s_o}\left[\frac{x_i}{s_x} - \frac{o_i}{s_o}\right] - \sum_i \alpha_i(x_i - o_i). \tag{273}$$

Let $W_o^a = 0$ and $o = u_T$, then

$$W^a(x) = \beta \ln(s_x/s_{u_T})\left(1 - \frac{||x - u_T||_2^2}{(s_x - s_{u_T})^2}\right) + \beta \sum_i \frac{x_i - u_{T,i}}{s_x - s_{u_T}}\left[\frac{x_i}{s_x} - \frac{u_{T,i}}{s_{u_T}}\right] - \sum_i \alpha_i(x_i - u_{T,i}) \tag{274}$$

Finding a closed-form solution for a global optimizer of (274) seems daunting not to mention the fact that the optimizer, $u_T$, appears in the optimization function. Fortunately, Theorem 4 states that a global optimizer of (274) is also a solution the corresponding VI($F, \mathcal{X} = [\epsilon, 1]^N$). If we assume $x = u_T$ lies in the interior of $\mathcal{X}$, then $F(u_T) = 0$.

$$F(x) = \beta \frac{1}{\sum_k x_k}(1 - \frac{x_i}{\sum_k x_k}) - \alpha_i = 0 \implies x_i = s_x(1 - \alpha_i \frac{s_x}{\beta}) \tag{275}$$

$$s_x = \sum_i x_i = s_x \sum_i (1 - \alpha_i \frac{s_x}{\beta}) \implies s_x\left[\frac{\sum_i \alpha_i}{\beta}s_x + (1 - N)\right] = 0 \implies s_x = \emptyset \text{ or } \frac{\beta(N-1)}{\sum_i \alpha_i} \tag{276}$$

Combining the two results, we find that the global optimizer is

$$u_{T,i} = \frac{\beta(N-1)}{\sum_k \alpha_k}\left(1 - \frac{\alpha_i}{\sum_k \alpha_k}(N-1)\right). \tag{277}$$

As the cost coefficients, $\alpha_i$, grow relative to the revenue coefficients, $\beta$, the optimal bid size drops. Also, as the number of users, $N$, increases, the optimal bid size increases, albeit with diminishing returns. Hence, *auto*-welfare has a rich dependence on the utility parameters and number of users, which is very different from the complete independence given by welfare.

The linear lower bound on *auto*-welfare regret is given by

$$regret_{W^a} = -regret_1 \geq \langle -F(x), x - u_T \rangle \tag{278}$$

$$= \sum_i \left[\beta\gamma_i \frac{x_i}{\sum_k x_k} - \alpha_i x_i\right] - \sum_i \left[\beta\gamma_i \frac{u_{T,i}}{\sum_k x_k} - \alpha_i u_{T,i}\right], \quad \gamma_i = (1 - \frac{x_i}{\sum_k x_k}) \tag{279}$$

$$= \sum_i u_i^\gamma(x) - \sum_i u_i^\gamma(u_T) \tag{280}$$

where $u_i^\gamma$ is the original utility function, $u_i$, with "revenues" ($\psi_i(M_i(x))$) reweighted by $\gamma_i$. In this case, *auto*-welfare regret is actually computing standard welfare with reweighted "revenues". High revenues are weighted lower and low revenues are weighted higher, which naturally encourages a more even distribution of resources.

**Congestion Control Protocols** Similar utility functions can be used to model a congestion control protocol with a *tail-drop* policy. In this game, a router drops packets if the total number of packets exceeds the network capacity (e.g., $C = 1$). In this case, the utility functions are defined piecewise:

$$u_i(x) = \begin{cases} x_i & \sum_k x_k \leq 1 \\ \beta \frac{x_i}{\sum_k x_k} - (\beta - 1) x_i & \sum_k x_k > 1 \end{cases} \tag{281}$$

For $\sum_k x_k \leq 1$, the utility functions are linear, thus the Jacobian of the associated game map, $F(x) = \{-\frac{\partial u_0}{\partial x_0}, \ldots, -\frac{\partial u_N}{\partial x_N}\}$, is a zero matrix, which is positive semidefinite. Monotonicity for the second case follows the same proof as for the linear resource allocation game above.

### A.11 Machine Learning Network Motivation

The example in the paper demonstrates our proposed no-regret algorithm on a cloud-based machine learning network. Our network is motivated by expectations of the next era of machine learning. Data is often the difference between a high performing model and a mediocre one; for some data hungry models (e.g., deep learning), *Big Data* launches them to state-of-the-art results. We expect *Big Data* to drive a mature digital supply chain capable of supporting an economy where producers provide data for consumers (i.e., machine learning models) to consume. Unlike the present, this commodity will not be transferred into local storage for consumption on personal machines; rather, it will be transmitted in batches, immediately consumed for training, and discarded to allow room for the next batch. Our model of a cloud-based machine learning network (MLN) is trivially adapted from the service oriented internet (SOI) model proposed in (Nagurney and Wolf 2014). In the original SOI model, service providers (e.g., Netflix, Amazon) stream content (e.g., movies, music). In our MLN model, service providers (e.g., Twitter, Wikipedia) stream machine learning data. Service providers control the quantity of data (i.e., # of samples × # of features) flowing through the market. Network providers charge service providers a fee for transmitting their data to consumers. The price different consumer markets are willing to pay service providers to stream data over a network of a certain quality is given by demand functions, price(quantity,quality). Given these relationships, service providers and network providers attempt to maximize their profits by varying their respective controls (quantity,quality) over the network. These relationships are parameterized so that we can instantiate ten five-firm networks by drawing parameters from uniform distributions over predefined ranges (code on github.com/*****).

### A.12 GTD Algorithms

The GTD update rules are

$$y_{t+1} = y_t + \alpha_t(b - A\theta_t - My_t) \tag{282}$$
$$\theta_{t+1} = \theta_t + \alpha_t(A^T y_t) \tag{283}$$

where $M = I$ or $M$ is a covariance matrix. Either way, $M$ is symmetric positive definite.

These update rules can be derived from a 2-player game with appropriate agent loss functions, $f^{(i)}$. For example, if

$$f^{(1)}(y, \theta) = -y^T(b - A\theta) + \frac{1}{2} y^T M y \tag{284}$$
$$f^{(2)}(y, \theta) = -\theta^T A^T y, \tag{285}$$

then OMoD with $F = [\nabla_y f^{(1)}, \nabla_\theta f^{(2)}]$ gives the same updates as GTD.

$$F = \begin{pmatrix} My + A\theta - b \\ -A^T y \end{pmatrix} \tag{286}$$
$$= \begin{pmatrix} M & A \\ -A^T & 0 \end{pmatrix} \begin{pmatrix} y \\ \theta \end{pmatrix} + \begin{pmatrix} -b \\ 0 \end{pmatrix} \tag{287}$$
$$= Jx + d \tag{288}$$
$$J_s = \begin{pmatrix} M & 0 \\ 0 & 0 \end{pmatrix} \tag{289}$$
$$z^T J_s z = \begin{pmatrix} \mathbf{z}_1^T & \mathbf{z}_2^T \end{pmatrix} \begin{pmatrix} M & 0 \\ 0 & 0 \end{pmatrix} \begin{pmatrix} \mathbf{z}_1 \\ \mathbf{z}_2 \end{pmatrix} \tag{290}$$
$$= \mathbf{z}_1^T M \mathbf{z}_1 \succeq \lambda_{\min} I \succ 0 \tag{291}$$

Therefore, $F$ is strongly monotone with parameter $\lambda_{\min}$.

By A.5, the corresponding path integral loss that these dynamics bound is

$$f(x) = \frac{1}{2}[x^T J^T x + x^T (J - J^T) o - o^T J^T o]$$
$$+ d^T (x - o) \quad (292)$$
$$= \frac{1}{2}(y^T M y - y_0^T M y_0) + y^T A \theta_0 - y_0^T A \theta - b^T (y - y_0) \quad (293)$$
$$= \frac{1}{2}\underbrace{(y^T M y - y_0^T M y_0)}_{y\text{-growth}} \quad (294)$$
$$- \underbrace{y^T (b - A\theta_0) + y_0^T (b - A\theta)}_{y\text{-alignment}} \quad (295)$$

In the GTD algorithms, $y$ was originally introduced as an auxiliary variable to estimate $\mathbb{E}[\rho \delta \phi] = b - A\theta$. If we let $\theta_0 = \theta$, and both $y$-growth $< 0$ and $y$-alignment $> 0$, this means that $y$ is a better estimate of $b - A\theta$ than $y_0$. This result is reassuring because it means that the GTD algorithm is naturally ensuring a bound on the error of this estimate, which is crucial to solving the "double sampling" problem. The GTD algorithm was originally developed with the goal of accurately estimating $b - A\theta$, however, the algorithm's development was motivated out of intuition. We hope this perspective provides a stronger mathematical explanation.

### A.13 Affine Wasserstein GANs

The minimax objective is

$$\min_G \max_d V(G, d) = \mathbb{E}_{x \sim p_{\text{data}}(x)}[d^T x] - \mathbb{E}_{z \sim p_z(z)}[d^T G z] \quad (296)$$

where $x \in R^n$, $z \in R^m$, $d \in R^n$, $G \in R^{n \times m}$.

For simplicity of exhibition, we will estimate the expectations with a single sample $x \sim p_{\text{data}}(x)$ and $z \sim p_z(z)$, however, the result applies to batches as well. The minimax objective simplifies to

$$\min_G \max_d V(G, d) = d^T (x - Gz) \quad (297)$$
$$= \sum_i d_i (x_i - \sum_j G_{ij} z_j). \quad (298)$$

Let $g = \begin{pmatrix} G_{11} \\ \vdots \\ G_{1m} \\ G_{i1} \\ \vdots \\ G_{im} \\ \vdots \end{pmatrix}$ be a flattened version of the matrix $G$. Also, let $A = \begin{pmatrix} z_1 & 0 & \cdots \\ z_j & 0 & \cdots \\ z_m & 0 & \cdots \\ 0 & z_1 & \cdots \\ 0 & z_j & \cdots \\ 0 & z_m & \cdots \\ \cdots & \cdots & \cdots \end{pmatrix} \in \mathbb{R}^{mn \times n}$. Then,

$$F = \begin{pmatrix} -d_1 z_1 \\ -d_1 z_j \\ -d_1 z_m \\ \vdots \\ -d_i z_j \\ \sum_j G_{1j} z_j - x_1 \\ \sum_j G_{ij} z_j - x_i \\ \sum_j G_{nj} z_j - x_n \end{pmatrix} \quad (299)$$

$$= \begin{pmatrix} 0^{mn \times mn} & -A \\ A^T & 0^{n \times n} \end{pmatrix} \begin{pmatrix} g \\ d \end{pmatrix} + \begin{pmatrix} 0 \\ -x \end{pmatrix} \quad (300)$$
$$= J\gamma + b \quad (301)$$
$$J_s = \begin{pmatrix} 0 & 0 \\ 0 & 0 \end{pmatrix} \succeq 0 \quad (302)$$

Therefore, $F$ is monotone. If $G$ and $d$ are regularized with $\alpha ||\cdot||_2^2$, then $F$ is strongly monotone with parameter $\alpha$.

By A.5, the corresponding path integral loss that these dynamics bound is

$$f(x) = \frac{1}{2}[x^T J^T x + x^T(J - J^T)o - o^T J^T o] + b^T(x - o) \tag{303}$$

$$= x^T J o + b^T(x - o) \tag{304}$$

$$= d^T(A^T g_0) - d_0^T(A^T g) - (d - d_0)^T x \tag{305}$$

$$= \underbrace{d^T(G_0 z)}_{G_0\text{-tricks-}d} - \underbrace{d_0^T(G z)}_{G\text{-tricks-}d_0} - \underbrace{(d - d_0)^T x}_{d\text{-realer-than-}d_0} \tag{306}$$

$$= V(d_0, G) - V(d, G_0). \tag{307}$$

If we let $d_0 = d$, then this loss is minimized if $G$ tricks $d$ better than $G_0$ tricks $d$; i.e., the generator, $G$, is an improvement over $G_0$. Alternatively, if we let $G_0 = G$, then this loss models how much the discriminator, $d$, has improved over $d_0$.

### A.14 Path Integral Loss for Minimax Games

$$\min_{x_1} \max_{x_2} V(x_1, x_2) \tag{308}$$

$$dV = \langle \frac{\partial V}{\partial x_1}, dx_1 \rangle + \langle \frac{\partial V}{\partial x_2}, dx_2 \rangle \tag{309}$$

$$f(x) = \int_{z:o \to x} \langle F(z), dz \rangle \tag{310}$$

$$= \int_{z:o \to x} \langle \frac{\partial V}{\partial z_1}, dz_1 \rangle + \langle -\frac{\partial V}{\partial z_2}, dz_2 \rangle \tag{311}$$

$$= \int_{\substack{z_1:o_1 \to x_1 \\ z_2:o_2 \to o_2}} dV - \int_{\substack{z_1:o_1 \to o_1 \\ z_2:o_2 \to x_2}} dV \tag{312}$$

$$= V(x_1, o_2) - V(o_1, o_2) - V(o_1, x_2) + V(o_1, o_2) \tag{313}$$

$$= V(x_1, o_2) - V(o_1, x_2) \tag{314}$$